\documentclass[referee]{aa} 
\usepackage{graphicx}
\usepackage[dvips]{color}
\usepackage{txfonts}

\begin{document}

  \title{Kepler Observations of Very Low-Mass Stars} 

  \author{E. L. Mart\'{\i}n \inst{1}\fnmsep\thanks{Visiting Astronomer, Kitt Peak National Observatory, National Optical Astronomy Observatory, which is operated by the Association of Universities for Research in Astronomy (AURA) under cooperative agreement with the National Science Foundation.}
 \and
J. Cabrera \inst{2}
\and
E. Martioli \inst{3,4}
 \and
           E. Solano \inst{5,6}
 \and
          R. Tata \inst{7}
          }

   \institute{Centro de Astrobiolog\'{\i}a (INTA-CSIC), Carretera de Ajalvir km 4, E-28550 Torrej{\'o}n de Ardoz, Madrid, Spain\
              \email{ege@cab.inta-csic.es}
       \and 
Institute of Planetary Research, German Aerospace Center, Rutherfordstrasse 2, D-12489, Berlin, Germany\
	\email{Juan.Cabrera@dlr.de}
      \and
Canada-France-Hawaii Telescope, 65-1238 Mamalahoa Hwy, Kamuela, Hawaii 96743, USA\
	\email{eder@cfht.hawaii.edu}
     \and
Laborat\'{o}rio Nacional de Astrof\'{i}sica (LNA/MCTI),  Rua Estados Unidos, 154 Itajub\'{a} - MG, Brazil\
	\email{emartioli@lna.br}
      \and
Centro de Astrobiolog\'{\i}a (INTA-CSIC), Departamento de Astrof\'{\i}sica,  P.O. Box 78, E-28691 Villanueva de la Ca\~{n}ada, Madrid, Spain\
              \email{esm@cab.inta-csic.es}
      \and
             Spanish Virtual Observatory\
      \and
Instituto de Astrof\'{\i}sica de Canarias, c/ V\'{\i}a L{\'a}ctea s/n, E-28200 La Laguna, Tenerife, Spain\
              \email{rrtata@iac.es}
\\
             }

   \date{Received 29 01, 2013; accepted 09 05, 2013}

 
  \abstract
  {Observations of very low-mass stars with Kepler represent an excellent opportunity to search for planetary transits and to characterize optical  photometric variability at the cool end of the stellar mass distribution. In this paper, we present low-resolution red optical spectra that allow us to identify 18 very low-mass stars that have Kepler light curves available in the public archive. Spectral types of these targets are found to lie in the range dM4.5--dM8.5, implying spectrophotometric distances from 17 pc to 80 pc. Limits to the presence of transiting planets are placed from modelling of the Kepler light curves. We find that the size of the planets detectable by Kepler around these small stars typically lie in the range 1 to 5 Earth radii within the habitable regions (P$\le$10 days). We identify one candidate transit with a period of 1.26 days that resembles the signal produced by a planet slightly smaller than the Moon. However, our pixel by pixel analysis of the Kepler data shows that the signal most likely arises from a background contaminating eclipsing binary. For 11 of these objects reliable photometric periods shorter than 7 days are derived, and are interpreted as rotational modulation of magnetic cool spots. For 3 objects we find possible photometric periods longer than 50 days that require confirmation. 
H$_\alpha$ emission measurements and flare rates are used as a proxies for chromospheric activity and transversal velocities are used as an indicator of  dynamical ages. These data allow us to discuss the relationship between magnetic activity and detectability of planetary transits around very low-mass stars. We show that Super-Earth planets with sizes around 2 Earth radii are detectable with Kepler around about two thirds of the stars in our sample, independently from their level of chromospheric activity.  
 }
  \maketitle
%

\section{Introduction}

Very low-mass (VLM) stars with spectral types between dM4 and dM7 constitute the most numerous population of the Milky Way (e.g., \cite{2012AA.542..A105L}, and references therein).  Besides their ubiquity, they are interesting in their own right 
because of interesting physical processes that take place among them.   
The transition between partially and fully convective stellar interiors 
is expected to take place at spectral subclass dwarf (d) M3.5 (\cite{2011MSAIt.in press}), and thus 
stars around this region are of considerable interest for studies of magnetic activity and rotation ({\cite{1998AA.331..581D};  \cite{2006ApJ.646..L73P}). 
In this work we consider the boundary between VLM stars and low-mass stars to be located at spectral subclass dM4  (M4 dwarf). 

The onset of metallic grain formation takes place at around spectral subclass dM7 (\cite{1997ApJ....480.L39J}).  Thus, dM7 objects and cooler are 
generically called ultracool dwarfs (UCDs). This term includes the L dwarfs (\cite{1997AA....327.L29M}) and cooler spectral types.  
Very late-M dwarfs and L dwarfs are known to display complex photometric variability that has been attributed to weather-like effects 
due to inhomogeneous distribution of dusty clouds in their photospheres (\cite{2001ApJ....557.822M} ; \cite{2002AA....389.963B}). 
On the other hand, magnetic cool spots are thought to be the dominant source of surface inhomogeneities in VLM stars (\cite{2004AA.421..259S}). 

The boundary between  VLM stars and brown dwarfs (BDs) is located at M7 spectral subclass 
 (\cite{1995Natur.377..129R}) in the Pleiades cluster. Most M dwarfs (dMs)  in the general field are older than 100 Myr   
(\cite{2010AA.517..53M}) and hence late-M dwarfs in the Kepler field are likely to have 
stellar masses. Nevertheless, the lithium test has shown that it is possible to encounter some young BDs of very late-M spectral 
subclass even in the immediate solar vicinity (\cite{1998MNRAS....296.L42T} ; \cite{1999AJ....118.1005M}). 

VLM stars have become attractive targets in the search for habitable rocky planets because their small masses and sizes  
favour the two most successful planet detection techniques, namely, high-precision radial velocity and transits, and renders 
transmission spectroscopy of Earth-sized planets in the habitable region around VLM stars potentially feasible with the future 
generation of large telescopes both in space as well as from the ground 
(\cite{2009IAUS..253...37I} ; \cite{2009ApJ.698..519K} ; \cite{2011A&A...525A..83B} ; \cite{2011A&A...529A...8R} ; \cite{2011ApJ...728...19P}).  

Radius measurements of a few VLM stars from angular diameter interferometric observations have recently become available 
(\cite{2001ApJ....551.L81L}). They have an accuracy smaller than 10\% and, combined with trigonometric parallaxes and photometry, have permitted the determination of empirical relations to convert astronomical observables to astrophysical parameters 
that agree fairly well with theoretical models. The coolest single star with a direct size measurement is GJ 551 (dM5.5) for which a radius of 0.141 $\pm$ 0.007 Rsun have been obtained (\cite{2009AA....505.205D}).  However, more accurate radii measurements using eclipsing binaries indicate that the theoretical models systematically underestimate the sizes of dM stars (\cite{2012ApJ.756.47S}). 

VLM stars represent a challenge for the search of planets because of their high levels of magnetic activity (e.g., 
\cite{2012MNRAS.427.3358G}; \cite{2012AJ....143...93R}), both from the point of view of transit searches and
characterization (\cite{2012AJ....143...12T}; \cite{2011AJ....141..166H}; \cite{2012AJ....144..145B};
\cite{2012ApJ...757..133L}) as from the point of view of radial velocity surveys (e.g., \cite{2012A&A...541A...9G}). 
Hence, future surveys for planets around VLM primaries are strongly motivated but their outcome may be optimized by appropriate 
selection of the best targets to minimize the effects of stellar activity. 

Using a combination of different surveys, it has been estimated that in a sample of 100 dMs there could be 1 transiting habitable super-Earth (\cite{2013AN.334..1B}).  The Kepler NASA space mission provides a unique opportunity to study the time variability of dMs and to search for transiting planets around them. In this work we consider a sample of 18 VLM stars, which is too small to expect the detection of 
one transiting planet, but it is nevertheless useful to estimate the sensitivity of Kepler to exoplanet transits at the low-mass tail 
of the stellar masses, and to study the effects of magnetic activity on planet detectability. 

The rest of this paper is organized as follows: Section 2 presents the selection criteria of the stars discussed in this paper. Section 3 
deals with the spectroscopic observations of the sample. Section 4 discusses the analysis of Kepler light curves. Section 5 covers the 
derivation of stellar parameters for the sample. Finally, Section 6 provides our final remarks regarding future planet searches in VLM stars.


\section{Target selection}

Our VLM star candidates were selected using information obtained from the Kepler Input Catalog (KIC), 
the Two Micron All Sky Survey (2MASS, \cite{2006AJ....131.1163S}), and the  Sloan Digital Sky Survey (SDSS, \cite{2000AJ....120.1579Y}). 
The following criteria were applied: 
\begin{itemize}
\item[$\bullet$] Xflg=0, Aflg=0, to avoid sources flagged in 2MASS as minor planets or contaminated by nearby extended sources.
\item[$\bullet$] 2MASS Kmag $>$9, to eliminate red bright sources likely to be giants. 
\item[$\bullet$] Kp $<$ 20 (KIC photometry), to get rid of sources too close to the Kepler confusion limit.
\item[$\bullet$] r $<$ 20 (SDSS photometry), to keep sources bright enough for optical spectroscopic follow-up.  
\item [$\bullet$] (r-J)$>$4.0 , J-K$<$1.05 (2MASS and SDSS) to select red sources with colors similar to late-M dwarfs 
(\cite{2011AJ.141.97W}) but not so red in the infrared that there could be 
contamination by red giants (\cite{1988PASP..100.1134B}).  
\item[$\bullet$] Total proper motions (PMs)  obtained from the KIC larger than 0.1 arcsec/year. 
\item[$\bullet$] Light curves publicly available in at least two Kepler quarters.  
\end{itemize}

The color cuts are shown in Figure 1 together with our VLM candidates and a subset of the stellar population in the Kepler field. 
Two additional KIC targets that do not meet all of the criteria given above were, nevertheless, included in our study; namely KIC 7435842, a photometric VLM star candidate without a PM value in the KIC, and KIC 8450707, a red giant candidate. As shown in the next section, our spectroscopic observations confirm KIC 7435842 as a VLM dwarf and KIC 8450707 as a red giant. 

\begin{figure}[t!]
    \label{Color}
    \centering
    \includegraphics[width=12.5cm]{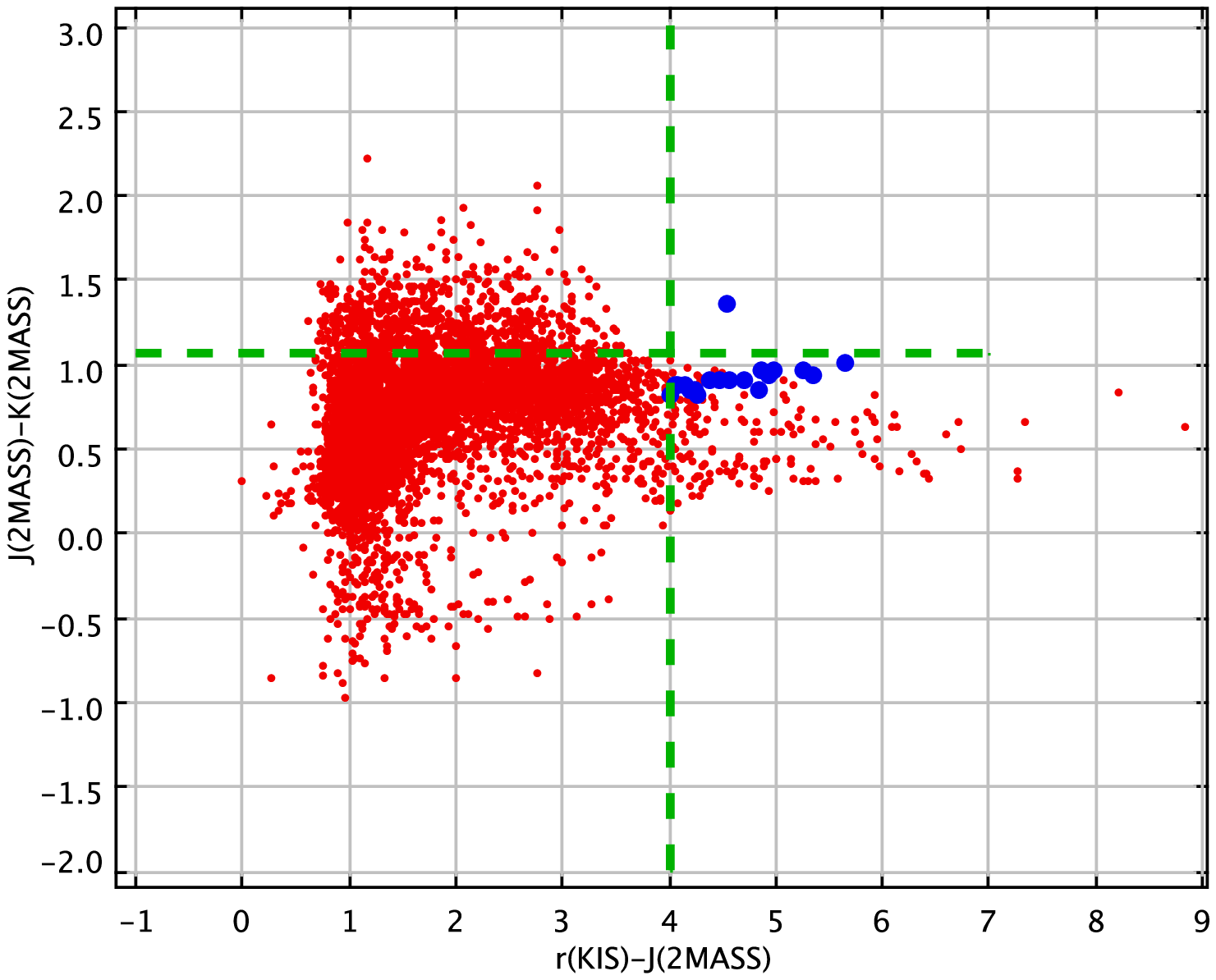}
    \caption{Color-color diagram showing our KIC targets (blue)  compared with a sample of 8000 KIC objects selected as described in the text but without color cuts. The only blue object outside the color cut boundaries is the giant.}
 \end{figure}

\begin{table*}
\centering
\caption{Astrometric and photometric data for our sample of VLM stars in the KIC.}
\label{tabla1}
\scriptsize
\hspace{-1cm}
\begin{tabular}{cccccccc}
\noalign{\smallskip}
\hline
\noalign{\smallskip}
KIC		& RA(J2000) &  Dec(J2000)         & r         &  Kp         & J            & J-K             & PM            \\
		& (hh:mm:ss.ss) & (dd:mm:ss.s)  &  (SDSS)  &  (KIC)   & (2MASS)   & (2MASS)       &  ($\arcsec$/year) \\
\noalign{\smallskip}
\hline
\noalign{\smallskip}
6751111   & 18 45 25.95 & 42 14 53.2  &  17.35 & 16.58 & 13.26 &  0.87  & 0.101   \\
7799941   & 18 45 43.65 & 43 31 08.5  &  19.52 & 19.53 & 13.86 &  1.00  & 0.116   \\
11597669 & 18 55 25.53 & 49 37 52.1  &  16.72 & 16.03 & 12.70 & 0.81  & 0.118  \\
8869922   & 19 00 05.65 & 45 07 06.1  &  16.03 & 16.05 & 11.89 & 0.87 & 0.134   \\
3219046   & 19 06 10.79 & 38 19 36.8  &  17.65 & 17.66 & 13.45 & 0.83  & 0.120   \\
4248433   & 19 09 37.68 & 39 23 26.0  &  15.72 & 15.74 & 11.33 & 0.89  & 0.132   \\
3330684   & 19 12 28.09 & 38 26 15.8  &  15.90 & 15.92 & 11.20 & 0.89  & 0.330   \\
5175854   & 19 13 30.91 & 40 22 26.9  &  18.30 & 18.32 & 13.83 & 0.90  & 0.174 \\
7517730   & 19 14 21.48 & 43 09 00.8  &  17.22 & 17.23 & 12.38 & 0.84  & 0.186 \\
7435842   & 19 15 59.81 & 43 05 41.6  &  17.94 & 17.95 & 13.00 & 0.92  &           \\
5353762   & 19 16 24.91 & 40 32 41.5  &  18.88 & 18.91 & 14.02 & 0.96  & 0.172 \\
12108566 & 19 20 23.52 & 50 37 16.2  &  17.27 & 16.53 & 13.21 & 0.87  & 0.143 \\
10538002 & 19 32 17.97 & 47 47 02.8  &  16.09 & 16.11 & 11.52 & 0.89  & 0.169 \\
7691437   & 19 38 48.26 & 43 21 22.8  &  18.06 & 18.07 & 12.71 & 0.92  & 0.284 \\
11356952 & 19 39 56.11 & 49 10 06.7  &  18.81 & 18.83 & 13.56 & 0.96  & 0.139  \\
9033543   & 19 43 07.79 & 45 18 09.8  &  16.31 & 16.33 & 11.33 & 0.95  & 0.303  \\
10285569 & 19 44 38.15 & 47 20 30.1  &  16.07 & 16.09 & 11.81 & 0.81  & 0.320  \\
8450707   & 19 52 44.62 & 44 28 51.6  &  14.24 & 13.51 &  9.70  & 1.35  &           \\
6233711   & 19 56 21.93 & 41 31 48.2  &  16.59 & 16.59 & 12.35  & 0.83 & 0.124 \\
\noalign{\smallskip}
\hline
\noalign{\smallskip}
\noalign{\smallskip}
\hline
\noalign{\smallskip}
\end{tabular}
\end{table*} 

Astrometric and photometric data for the VLM candidates can be found in Table 1. Coordinates, Kepler passband magnitudes and 
proper motions were taken from the KIC. r-band magnitudes were drawn from the SDSS, and near-infrared magnitudes were obtained from 
2MASS. 

Only one of our targets has been studied before, namely, KIC 3330684, which is also named LSPM J1912+3826. It was identified as a star with annual proper motion higher than 0.15 arcsec (\cite{2005AJ....129.1483L}). It is included in the stellar variability analysis of a sample of 129,000 dwarfs in the early data release of Kepler (\cite{2011AJ....141.108C}).

\section{Spectroscopic follow-up}

In order to assess the nature of  VLM candidates, low-resolution red optical spectroscopy is well-known to be an effective method. Thus, follow-up spectroscopic observations of the targets were carried out using three different telescopes as follows: 
 
\begin{itemize}
\item[$\bullet$] On 1 September 2011, and on 15-17 July 2012, the 4-meter Mayall telescope at Kitt Peak National Observatory (KPNO) with the R.C./Cryocam spectrograph was used to obtain reconaissance spectra of the bulk of our VLM KIC candidates. The sky was patched with clouds, and often during the nights the dome had to be closed due to thunderstorms. The BL181 grating and the 1.0 arcsec slit gave an effective resolution of 4.7 \AA  (FWHM=1.69 pixels) as measured on emission lines.  Order-sorting filters OG530 and GG475 were used to cutoff second-order light in the 2011 and the 2012 runs, respectively. The spectra have adequate signal to noise ratio to be usable in the spectral range from 600 to 920 nm.  The spectral type standard GJ905 was observed in 2011 for comparison purposes. 
\item[$\bullet$]  On 2 October 2011, the 2.5-meter Nordic Optical Telescope (NOT) with ALFOSC in the long slit observing mode was used to obtain spectra of KIC 7799941 and of KIC  11356952. The sky was clear and the seeing was below 1 arcsec. The grism number 5 and the 1.0 arcsec slit gave an effective resolution of 10.5 \AA  (FWHM=2.24 pixels) as measured on emission lines.   Order-sorting filter OG515 was used to cutoff second-order light. 
\item[$\bullet$]  On 10 July 2012, the 10.4-meter Gran Telescopio de Canarias (GTC) with OSIRIS in the long slit observing mode was used to obtain a spectrum of KIC 5353762. The sky was clear and the seeing was below 1 arcsec. The R500R grating and the 1.0 arcsec slit gave an effective resolution of 10.5 \AA (FWHM=2.24 pixels) as measured on emission lines.  
\end{itemize}
 
All of the spectra were debiased, flatfielded and extracted using the IRAF twodspec tools. Wavelength calibration were performed using 
arc lamps observed each night. The flux standard star EGGR39 was observed for calibration in every run and it was used to correct for 
instrumental response and telluric contamination. 
Table 2 provides additional details of the observations of each target as well as the main spectroscopic results which are described next. 
Figure 2 displays all of our final spectra, and shows the main spectral features used in the analysis.  

\begin{figure}[t!]
    \label{Spectra}
    \centering
    \includegraphics[width=13.5cm]{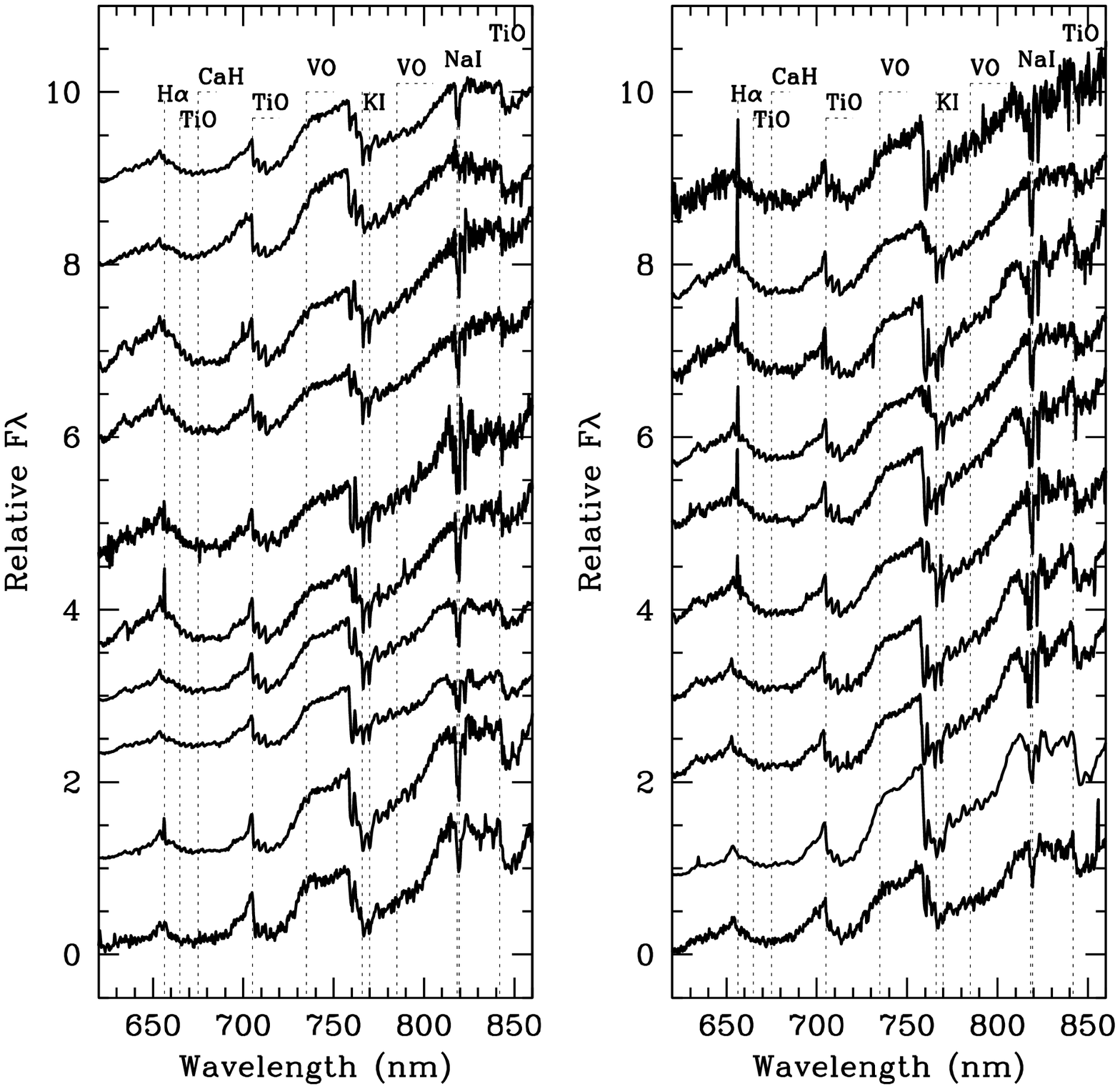}
    \caption{Final spectra for our 18 confirmed KIC VLM dwarfs and one spectral type reference star (GJ 905) and a giant (KIC 8450707). 
Left panel, from top to bottom the objects are: GJ 905 (dM5), KIC 8450707 (M4III), KIC 6233711 (dM4.5), KIC 10285569 (dM4.5), 
KIC 3219046 (dM4.5), KIC 6751111 (dM4.5), KIC 3330684 (dM5), KIC 9033543 (dM5.5), KIC 7691437 (dM7.5) and KIC 11356952 (dM8.5). The main spectral features are labelled. Right panel, from top to bottom the objects are: KIC 11597669 (dM5), KIC 8869922 (dM5), KIC 5175854 (dM5), 
KIC 4248433 (dM5), KIC 12108566 (dM5), KIC 10538002 (dM5), KIC 7517730 (dM6), KIC 7435842 (dM6), KIC 5353762 (dM6.5), KIC 7799941 (dM7).}
 \end{figure}

\subsection{Spectral typing and equivalent width measurements}

The low-resolution spectra of VLM candidates were compared with reference spectra of late-M dwarfs coming from the literature 
(\cite{1994AJ....108.1437H}; \cite{1999AJ....118.2466M}). Spectral types were derived by visual inspection of the closest spectral match. 
At these late-M spectral types the spectral regions around the VO bandhead at 730 nm and the TiO bandhead around 840 nm 
are particularly useful for classification, as well as the steep pseudocontinuum slope between 760 and 810 nm.  

\begin{table*}
\centering
\caption{Spectroscopic log and results.}
\label{tabla2}
\scriptsize
\hspace{-1cm}
\begin{tabular}{cccccccc}
\noalign{\smallskip}
\hline
\noalign{\smallskip}
KIC	& Date & Tel.  & Texp & Airmass & EW (H$_\alpha$) & EW (NaI) & SpT \\
 	&  (UT)   &         & (s)       &               & (\AA )             & (\AA )       &         \\
\noalign{\smallskip}
\hline
\noalign{\smallskip}
6751111   & 17 July 2012 & KPNO  &  900     & 1.23   &   -5.9$\pm$0.3   &  8.3$\pm$0.2  &  dM4.5      \\
7799941   &   2 Oct 2011  & NOT   & 1200    &  1.97  &   $>$-0.8            &  8.6$\pm$0.6  &  dM7         \\ 
11597669 & 17 July 2012 & KPNO  &  360     & 1.60  &   -8.4$\pm$0.9   &  8.8$\pm$0.8  &  dM5         \\
8869922   & 17 July 2012 & KPNO  &  600     & 1.03  &  -10.1$\pm$0.3  &  7.4$\pm$0.3  & dM5          \\
3219046   & 16 July 2012 & KPNO  &  600     & 1.01  &   -3.6$\pm$0.4   &  7.8$\pm$0.6  &  dM4.5      \\
4248433   & 17 July 2012 & KPNO  &  300     & 1.03  &   -6.9$\pm$0.2   &  7.0$\pm$0.2  &  dM5         \\
3330684   & 15 July 2012 & KPNO  &  600     & 1.28  &   -0.9$\pm$0.2   &  7.4$\pm$0.4  &  dM5        \\
5175854   & 17 July 2012 & KPNO  &  1200   & 1.41  &   -7.3$\pm$0.3   &  8.7$\pm$0.3  &  dM5        \\
7517730   & 16 July 2012 & KPNO  &  600     & 1.11  &   -1.2$\pm$0.2   &  8.2$\pm$0.2  &  dM6        \\
7435842   & 15 July 2012 & KPNO  &  1800   & 1.05  &   -1.8$\pm$0.2   &  9.1$\pm$0.2  &  dM6       \\
5353762   & 10 July 2012 & GTC    &  360     &  1.10 & $>$-1.0              &  6.6$\pm$0.3   &  dM6.5       \\
12108566 & 17 July 2012 & KPNO  &  600     & 1.28  &    -9.1$\pm$0.4  &  7.7$\pm$0.3  &  dM5       \\
10538002 & 17 July 2012 & KPNO  &  300     & 1.05  &  -4.5$\pm$0.2   &  7.9$\pm$0.3   &  dM5       \\
7691437   &   1 Sept 2011 & KPNO & 1800    & 1.58  & -10.2$\pm$0.3  & 10.6$\pm$0.3  &  dM7.5     \\
7691437   & 15 July 2012 & KPNO  &  900     & 1.16  & -8.4$\pm$0.2    & 10.3$\pm$0.2  &  dM7.5     \\
11356952 &    2 Oct 2011 & NOT   &  900     & 1.74  &   -5.8$\pm$0.5 &    8.3$\pm$0.8  &  dM8.5        \\ 
9033543   &  1 Sept 2011 & KPNO  &  600     & 1.03  & $>$-0.7               &   7.8$\pm$0.2  &  dM5.5         \\
10285569 & 16 July 2012 & KPNO  &  300     & 1.04  & $>$-0.6               &   7.7$\pm$0.5  &  dM4.5      \\
8450707   & 15 July 2012 & KPNO  &  300     & 1.08  & $>$-0.8             &  3.4$\pm$0.5   &  M4 III    \\
6233711   & 17 July 2012 & KPNO  &  900     & 1.06  & -1.6$\pm$0.2    &  7.2$\pm$0.2   &  dM4.5       \\
GJ 905       &   1 Sept 2011 & KPNO & 120     & 1.04  & -1.5$\pm$0.2     & 7.9$\pm$0.2    &  dM5 std.   \\
\noalign{\smallskip}
\hline
\noalign{\smallskip}
\noalign{\smallskip}
\hline
\noalign{\smallskip}
\end{tabular}
\end{table*} 

Equivalent widths of H$_\alpha$ in emission, a chromospheric activity and mass accretion indicator (e.g., \cite{2003AJ....126.2997B}), were measured by gaussian fitting using the IRAF task splot. If no emission was seen, an upper limit was derived by direct integration of a 20 \AA  
~region centered on the wavelength of the H$_\alpha$ line (656.2 nm). None of the targets showed an H$_\alpha$ emission strong enough to indicate active mass accretion. 

Equivalent widths of the NaI subordinate doublet at 818,819 nm, an indicator of surface gravity in field late-M dwarfs (\cite{2010AA.517..53M}),  were measured by direct integration with splot. All the equivalent width measurements are given in Table 2. 
None of the targets was found to have any spectroscopic indication of extreme youth (age $\le$100 Myr), such as NaI absorption equivalent width weaker than 6 \AA , which is typical of Pleiades members of similar spectral subclass (\cite{1995MNRAS.272..630S}).
Thus, our spectroscopic results confirm the identification of 18 mature VLM stars in the Kepler public database.

\section{Kepler light curves}

All of the Kepler publicly released light curves for our targets have 
been utilized for this work. 
Most targets were observed in four quarters (Q6, Q7, Q8 and Q9) for
GO20001 and GO20031. Two targets, namely KIC 3330684 and KIC 9033543, have been observed 
continuously since Q1. The baseline in days and number of usable light curve points for each
target were computed from the FITS header keywords NAXIS2 and EXPOSURE, and are summarized in Table 3. 
All of the targets were observed with the long cadence mode, and hence
each point of the light curve corresponds to an exposure time of 30
minutes. 

The light curves and the pixel mask data of our targets were downloaded by ftp from the Kepler mission
archive at MAST in FITS format.  These data files were used to search for transits, 
and to determine periodicities and flare duty cycles as described below.  

\begin{table*}
\centering
\caption{\label{t3} Kepler light curve results.}
\label{tabla3}
\scriptsize
\hspace{-1cm}
\begin{tabular}{cccccccc}
\noalign{\smallskip}
\hline
\noalign{\smallskip}
KIC	& Np         & Baseline & Period & Flare number &   Flare rate   & Precision (5 $\sigma$)  & Rp/R$_Earth$   \\
 	&  (points)  & (days)    & (days) &                      &                     & (\%)                               &                          \\
\noalign{\smallskip}
\hline
\noalign{\smallskip}
6751111   &  16820  &  316.4  & 0.37,0.47  & 10       &   0.0023           &  2.0   &   2.6          \\ 
7799941   &  16820  &  316.4  &  53:           &   0       &   $<$0.0001    &  8.6   &   3.2          \\ 
11597669  &  13541 &  254.7  &  0.76         &   6       &   0.0022          &  2.7   &   2.7          \\ 
8869922   &  16820  &  316.4  &  0.70         & 14       &   0.0043          &  1.4   &   2.0(*)       \\ 
3219046   &  16820  &  316.4  &  0.49         &  4        &   0.0009          &  4.0   &   3.7           \\
4248433   &  16820  &  316.4  &  0.42         &  1        &   0.0002          &  9.5   &   5.4(*)        \\
3330684   &  54006  & 1016.0 & 51:            &  5        &   0.0005          &  0.2   &   0.8           \\
5175854   &  16820  &  316.4  &  0.56         &  0        &   $<$0.0001   &  4.7   &   3.8           \\
7517730   &  16820  &  316.4  & 62:            &  1        &   0.0002         &  0.7   &   1.2           \\
7435842   &  16820  &  316.4  &                  &  0        &   $<$0.0001   &  1.3   &   1.6           \\
5353762   &  16820  &  316.4  &  33:           &  0        &   $<$0.0001   &  3.6   &   2.6             \\
12108566  &  16820  &  316.4  & 2.79        &  54       &  0.0154            &  3.2   &   3.1(*)           \\
10538002  &  16820  &  316.4  & 1.94        &  38       & 0.0102            &  0.8   &   1.6(*)          \\
7691437   &  16820  &  316.4  & 0.88         &   0        & $<$0.0001     &  2.6   &   2.6(*)           \\
11356952  &  16820  &  316.4  &  0.16       &  19       &  0.0044          &  5.2   &   2.0               \\      
9033543   &  54006  & 1016.0  &  50:         & 28        &  0.0020          &  0.3   &   0.9                \\  
10285569  &  12445  &  234.1  &                &   0        &  $<$0.0001   &  0.3   &   1.0                \\
8450707   &    476  &    9.0       &                &  0         &  $<$0.0001   &   -    &    -                   \\ 
6233711   &  16820  &  316.4  &  6.53        &  3         &  0.0005         &  0.4   &   1.2             \\  
\noalign{\smallskip}
\hline
\noalign{\smallskip}
\noalign{\smallskip}
\hline
\noalign{\smallskip}
\end{tabular}
\end{table*}

\subsection{Planet transit search} 

We have used the algorithm DST (\cite{2012A&A...548A..44C}) for the search
of transiting planets in the Kepler light curves of our targets. 
The algorithm proceeds in two steps; first it removes the long term
stellar variability with a Savitzky-Golay algorithm and the periodic
stellar variability with a harmonic filter at the frequencies found
by a Lomb-Scargle analysis of the data.
The second step searches for the periodic signature of transiting
planets. We have not found the signature of any reliable planetary transiting
signal, although we have found transit-like features in the light
curves of 3330684, 7517730, and 9033543, which are discussed below.

For this study, we have established two type of constraints, the first
one based on the expected performance of transit detection algorithms
and the second one, more conservative, based on the analysis of the 
scatter of the filtered light curves.

The first constraint follows the procedure used by \cite{2012ApJS..201...15H}. 
Assuming that the performance of transit detection algorithms evolves
as expected for photometric surveys (\cite{2006MNRAS.373..231P}), and considering the 
detection performance of Kepler as studied by \cite{2012ApJS...204.24B} 
one can estimate the minimum planetary size detectable in our sample at a 
given stellar magnitude. 12 out of the 18 dwarfs considered in this study have Kepler magnitudes brighter than 17.5, and, 
extrapolating the performance of Kepler, we expect to be able to detect around these stars transit signals with
depths of 500 ppm or larger. For a star of 0.15 solar radii, this means a
planet of the size of 0.37 Earth radii (only 30\% larger than the
Moon), which would be a breakthrough for transiting planets
(Kepler-42d has a size of 0.57 Earth radii, the size of Mars, see
\cite{2012ApJ...747..144M}). 

Transit detection algorithms detect regularly
periodic signals with amplitudes below  the 1$\sigma$ photometric scatter which are 
undetectable by ocular inspection of individual transits. 
We have chosen, however, to set more conservative planet detections limits because we are aware that
the stellar activity filtering and transit detection tools used in {\bf our} study were optimized for solar-like stars.
The stars in our sample show very irregular activity patterns, in
particular flares and variable rotational modulation (in terms of amplitude and phase),  
which are challenging to filter out. Therefore, we have used a second constraint, much more conservative,
in the same way as it is done for the search of transits for planets
detected by radial velocity (\cite{2012ApJ...761...46W}). 

The principle of our conservative approach is to study the scatter of the light curve to rule out 
single transit features with a certain confidence level.
The main difficulty  is that transits of planets
around VLM dwarfs last typically less than 1 hour, which is comparable with
the Kepler long cadence integration time, and this means that single transit events
might be represented by very few (2-3) measurements.
 If the residuals were normally distributed, an event that deviates 5
$\sigma$ from the mean would be produced only once in every 1.74
million events.
Typically, we have around 16,000 measurements per light curve (Table 3), so one
would expect a very low impact from such outliers (typically, $<$1 case in 100 light curves). 
However, we must be aware that the outliers may not be normally distributed, and
that there could be residual instrumental effects in the data which can
mimic a transit (see discussion below). 
Therefore, for our conservative approach, we have tried to rule out
only periodic transits, not single events. 

We have chosen periods comparable to the expected position of the
habitable zone of the host star (P$\le$10 days; \cite{2009ApJ.698..519K}) and ruled 
out transits with a depth of 5 times the scatter (5 $\sigma$) by inserting transits of artificial planets  
in the light curves. We checked that the DST algorithm was able to detect all of the simulated transits down to 
the sensitivity limits provided in Table 3, where we give the 5 $\sigma$ scatter
of the filtered light curve (our photometric precision) and the radius of the
planet ruled out by our study in Earth units. For active stars (those denoted with asterisks in the 
eighth column of Table 3), we adopted a 10 $\sigma$ precision limit to estimate the radii of the planets. 
Typical values of planets detectable with DST in this sample range from 1 to 5 Earth radii.  
The dependence of the detectability of planet sizes with respect to Kepler magnitude is shown in 
Figure 3. The planet size detectability does not depend linearly with target brightness because fainter 
stars tend to be cooler and have smaller radii than larger stars, and this compensates for their faintness. 

\begin{figure}[t!]
    \label{figKpPs}
    \centering
    \includegraphics[width=12.5cm]{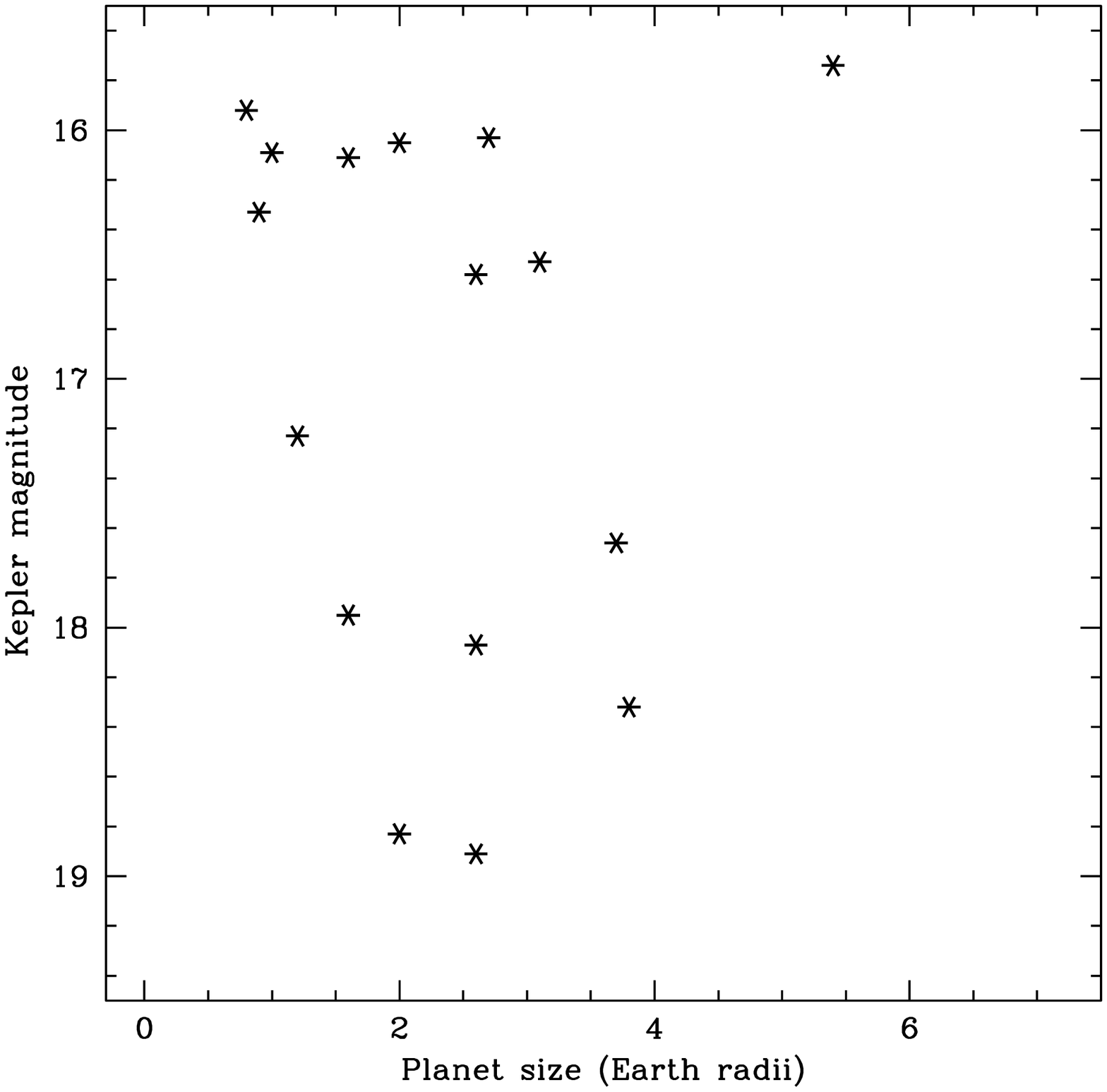}
    \caption{Detectability of planets of different sizes with respect to Kepler magnitude in our sample of VLM stars. Kepler magnitudes are 
listed in Table 1. Planet sizes are given in units of Earth radii and are listed in Table 3.}
 \end{figure}

\subsubsection{The case of KIC 3330684} 

As an example of the short-period transit signals which can be detected by Kepler in VLM stars,  
we focus on KIC 3330684.  
The DST analysis, when applied to the whole public Kepler
data set consisting of 13 quarters (about $1\,141$ days),
shows the presence of a periodic transit-like signal with a period of
$1.261\,319\,8 \pm 0.000\,001\,6$ days, an epoch $2\,451\,677.431\,09
\pm 0.000\,95$ in Julian Date, a depth of $190 \pm 11$ ppm and a
duration of $1.4 \pm 0.1$ h (Figure 4). 
The detection of this signal is compatible with the detection limit
expected from the candidate distribution of \cite{2012ApJS...204.24B}, 
and it would correspond, for a central transit and considering the radius
of the dM5 host, to an object of 0.24 Earth radii or 0.9 times the
radius of the Moon.
However, the expected duration for a central transit around a dM5 star
with this orbital period, and considering a circular orbit, is only 45
minutes, incompatible with the ephemeris found for the periodic
signal.  An eccentric orbit could account for a longer transit duration (see,
for example, the case of Gliese 876 d, \cite{2010ApJ...719..890R}),
but the most simple explanation is the presence of a contaminating
eclipsing binary (CEB). 

\begin{figure}[t!]
    \label{KIC3330684_CEB}
    \centering
    \includegraphics[width=12.5cm]{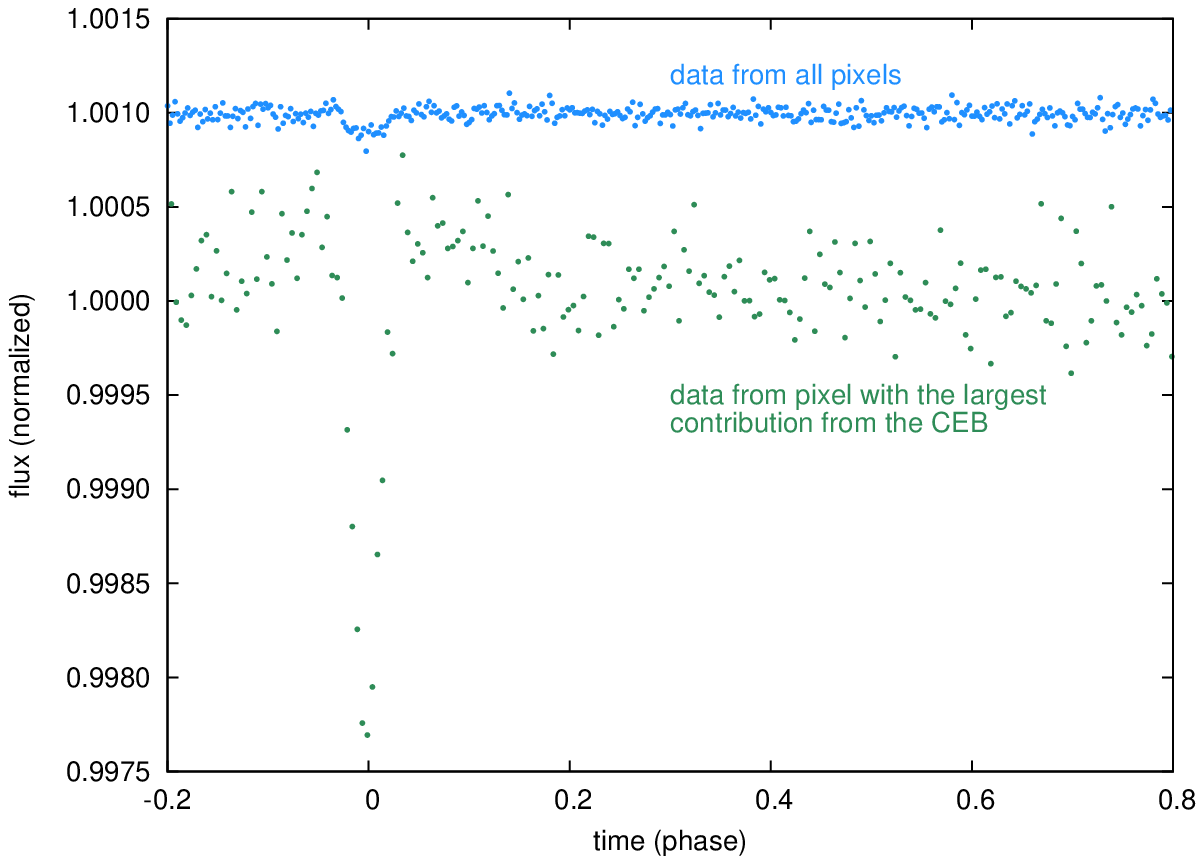}
    \caption{Periodic transit-like signal in the light curve of KIC 3330684 folded with a period of $1.261\,319\,8 \pm 0.000\,001\,6$ days. Our pixel by pixel analysis indicates that the signal comes from an unidentified CEB located to the SW of the target.}
 \end{figure}     

We have searched for the CEB responsible for the periodic signal using
the centroid motion of the light curve and the information of the
individual pixels of the Kepler mask (Figure 5), following the approach of
\cite{2010ApJ...713L.103B}.  Their method is the analysis of the centroid motion and the 
study of the behaviour of the individual pixels of the Kepler 
mask.  
However, the information obtained during a single quarter is not
enough to prove the presence of a CEB because the SNR is extremely
low. By studying the position of the Kepler pixels in the sky, we have
combined the light of the pixels corresponding to the same region of
the sky as observed in the different quarters. 
We have found that the $\sim 200$ ppm (0.02\%) signal found by DST
originates somewhere to the SE of the main target and has a depth
about 10 times larger (0.2\%) and the clear V-shape of the grazing
eclipse of a stellar binary (Figure 4). 
We conclude that the origin of the periodic signal is not on target,
although we have not been able to unambiguously identify, based on the
available dataset, the nature of the CEB. We have not found any source at 
the location of the CEB in the KIS 
catalog down to a magnitude limit of 21.5 in the IPHAS r-band (\cite{2012AJ...144..24G}). 
However, this limit may not be sensitive enough. In order to produce the observed 
dips of 0.02\% depth, the CEB can be up to 9.2 mag. fainter than KIC 3330684. 
Given that the KIS r-band magnitude of KIC 3330684 is 15.82, deep imaging down 
to a sensitivity of 25th mag. in the optical is required to identify the CEB.  

\begin{figure}[t!]
    \label{KIC3330684_pixel}
    \centering
    \includegraphics[width=11.7cm]{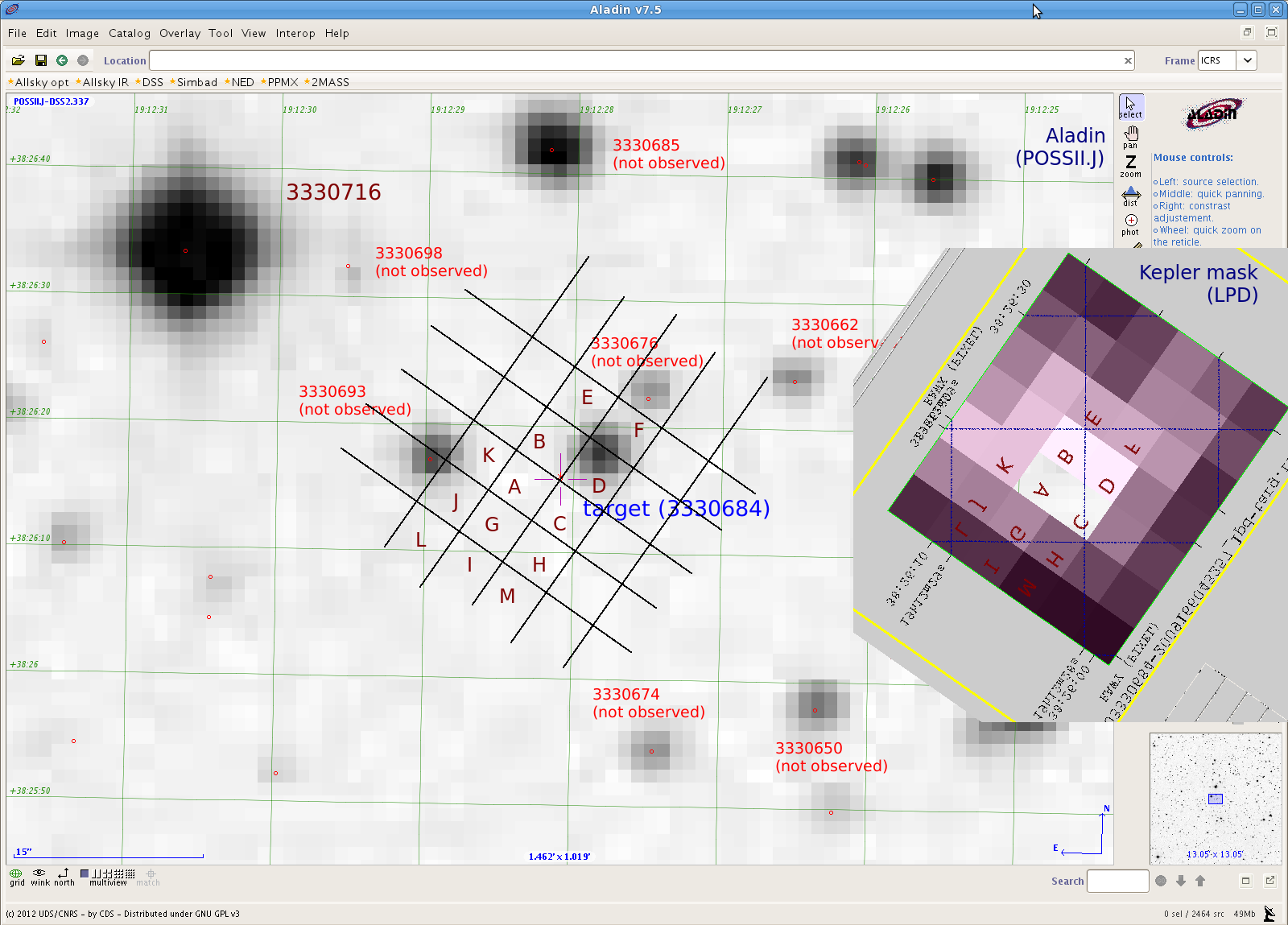}
    \caption{Digital Sky Survey image of the field around KIC 3330684 with the Kepler pixel mask superimposed. The Kepler pixels analyzed in 
our study are labelled with capital letters.}
 \end{figure}     

This example shows that we are able to detect extremely shallow
short-period signals in the Kepler light curves of a VLM star. 
Finally, we note that KIC 3330684 is neither in the list of planetary
candidates, nor in the list of false alarms of Kepler objects of
interest (\cite{2012ApJS...204.24B}), however it is cited as a Kepler
threshold crossing event by \cite{2012ApJS...199.24T}.

\subsubsection{KIC 7517730 and KIC 9033543} 

As an example of instrumental residuals that can mimic transit-like
events, we mention the examples of KIC 7517730 and KIC 9033543.
KIC 7517730 shows two events with depths of around 1\% and 2\%
(corresponding to planetary sizes, for central transits, of 1.4 and 2
Earth radii, respectively) as illustrated in Figure 6. 
However, the shape of those events is similar to the pattern produced
by instrumental effects identified in Kepler data, such as temperature
fluctuations of the instrument or perturbations produced by solar
particles (see \cite{KSCI-19040-003}).
When comparing the Kepler single aperture photometry (SAP) and the
pre-search data conditioning photometry (PDCSAP), one can see how the
PDCSAP correction enhances these features, which is a sign of their 
non-astrophysical origin. SAP is the flux in units of electrons per second contained 
in the optimal aperture pixels collected by the spacecraft. 
PDCSAP is the flux contained in the optimal aperture in 
electrons per second after the PDC (Pre-search Data Conditioning) 
module has applied its detrending algorithm to the PA (Photometric 
Analysis module) light curve. The Kepler data files are described in the document 
"Kepler archive manual" KDMC-10008-004
(http://keplergo.arc.nasa.gov/Documentation.shtml). 

\begin{figure}[t!]
    \label{KIC75177_cand}
    \centering
    \includegraphics[width=12.5cm]{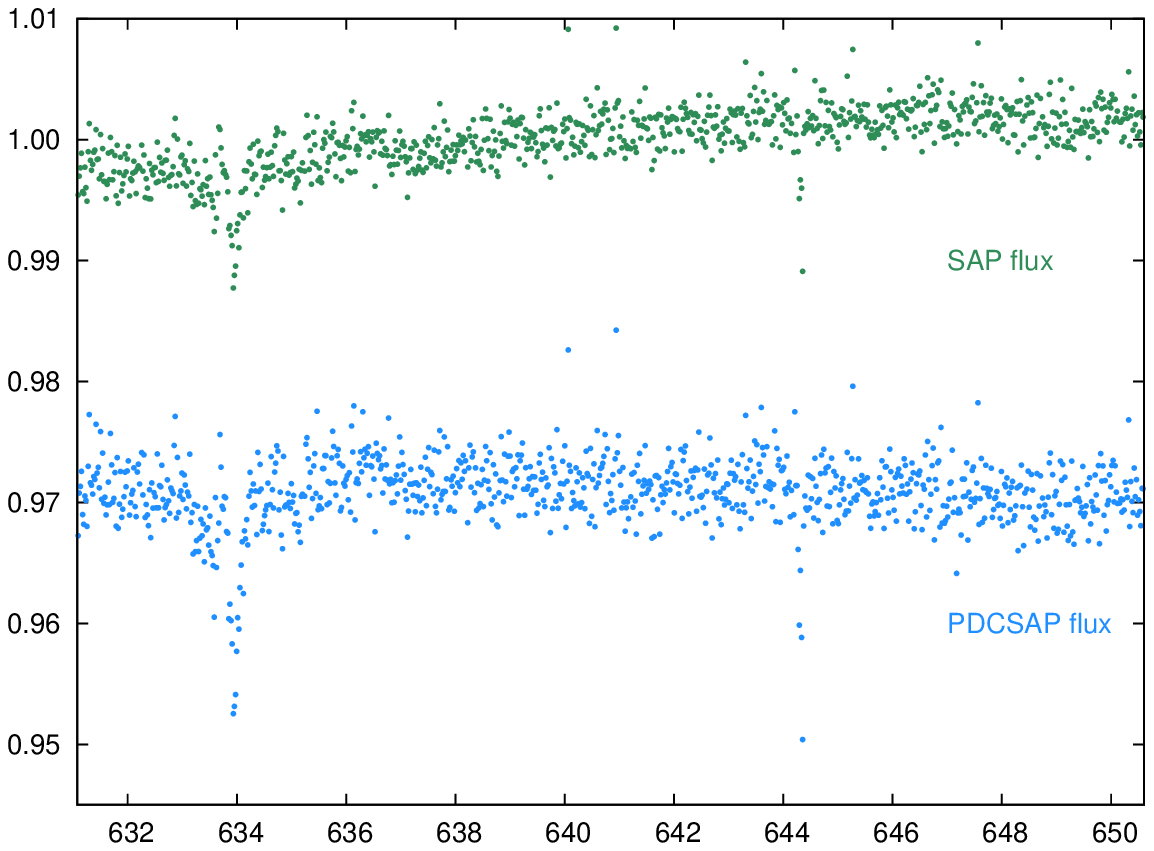}
    \caption{Transit-like signals in the light curve of KIC 7517730 which are likely due to instrumental effects as discussed in the text.}
 \end{figure}    

KIC 9033543 shows an odd pattern which we have not been able
to find in the available documentation. 
It mimics the transit of a multiple system, and therefore it is
extremely interesting, because multiple systems are thought to be
bona-fide planetary systems (\cite{2012ApJ...750..112L}).
However, in this case the shape and the depth of the events depend on
the correction applied to the data, and therefore is indicative of an
instrumental origin (Figure 7).

\begin{figure}[t!]
    \label{KIC9033_cand}
    \centering
    \includegraphics[width=12.5cm]{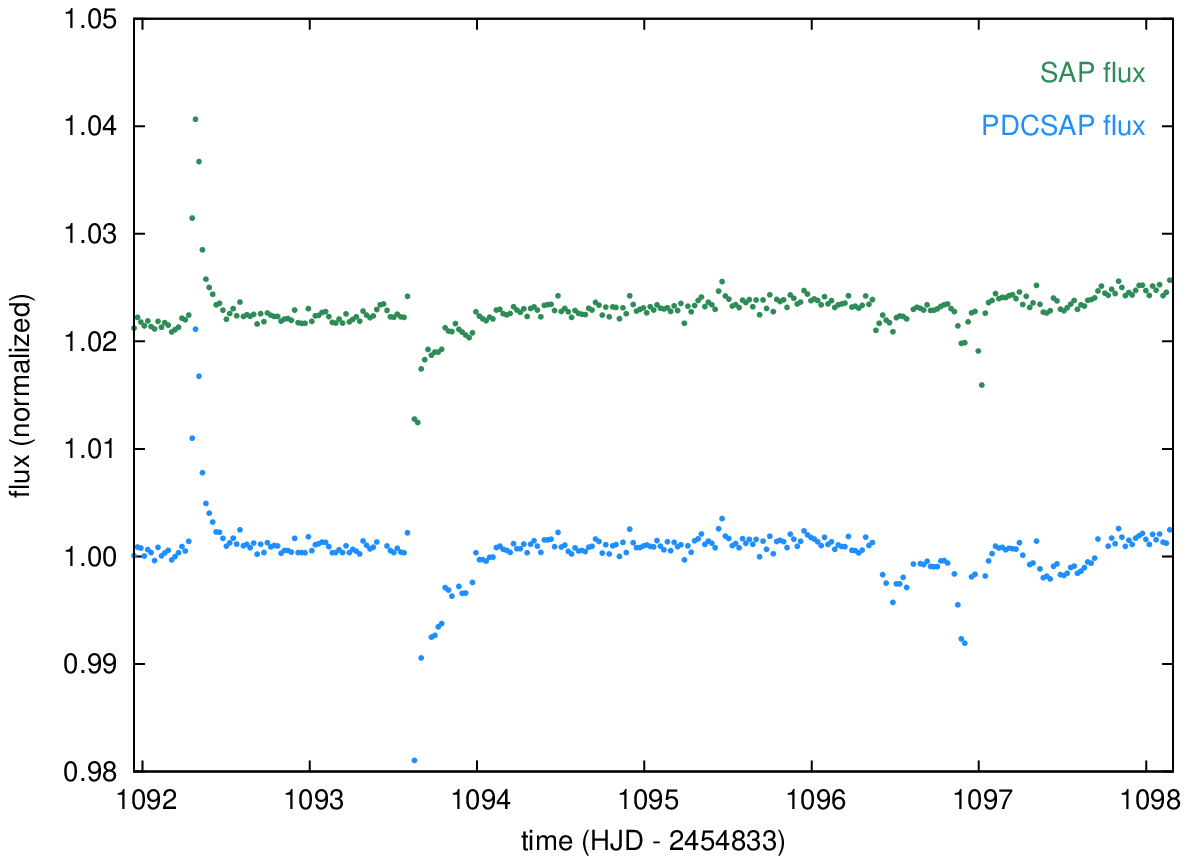}
    \caption{Transit-like signals in the light curve of KIC 9033543 which are likely due to instrumental effects as discussed in the text.}
 \end{figure}     
 
These two examples illustrate the difficulty of characterizing the
astrophysical origin of single transit events and advocate for the use
of the conservative constraints adopted in this paper.

\subsection{Stellar variability}

\subsubsection{Rotational periods}

Rotational periods were  determined as a by-product of the
planetary transit search described above. 
As an independent check, the Kepler light curves of the targets were
also fed into NASA Star and Exoplanet Database (NStED) periodogram
service. By default this service uses the Lomb-Scargle algorithm, but the
box-fitting least squares and the Plavchan algorithms are also
available. We used these three algorithms to check the robustness of the periods found by 
DST. Each quarter of data was analyzed independently. The periods provided in Table 3 were confirmed with 
two or more different algorithms. The periods marked with colons were not confirmed by two or more different 
algorithms, and hence they are considered as more uncertain. All of the confirmed periods have a low false alarm probability 
($<$ 0.01, as defined by \cite{1982ApJ...263.835S}). We interpret all these periodicities as due to 
rotational modulation of the light curves cause by spots in the atmospheres of the dwarfs. 

KIC 6751111 is the only target in our sample that showed two
persistent periodicities at 0.37 and 0.47 days (Figure 8). 
This anomaly suggests that this object could be an unresolved binary
composed of two components of similar brightness. It may be a member 
of the growing class of detached short-period M-dwarf binaries recently discussed by 
\cite{2012MNRAS...425.950N}. 
The Kepler light curve is well fitted by our algorithm using a combination 
of two sinusoidal curves with the periods derived from the periodogram (Figure 9). 

\begin{figure}[t!]
    \label{KIC6751111_P}
    \centering
    \includegraphics[width=12.5cm]{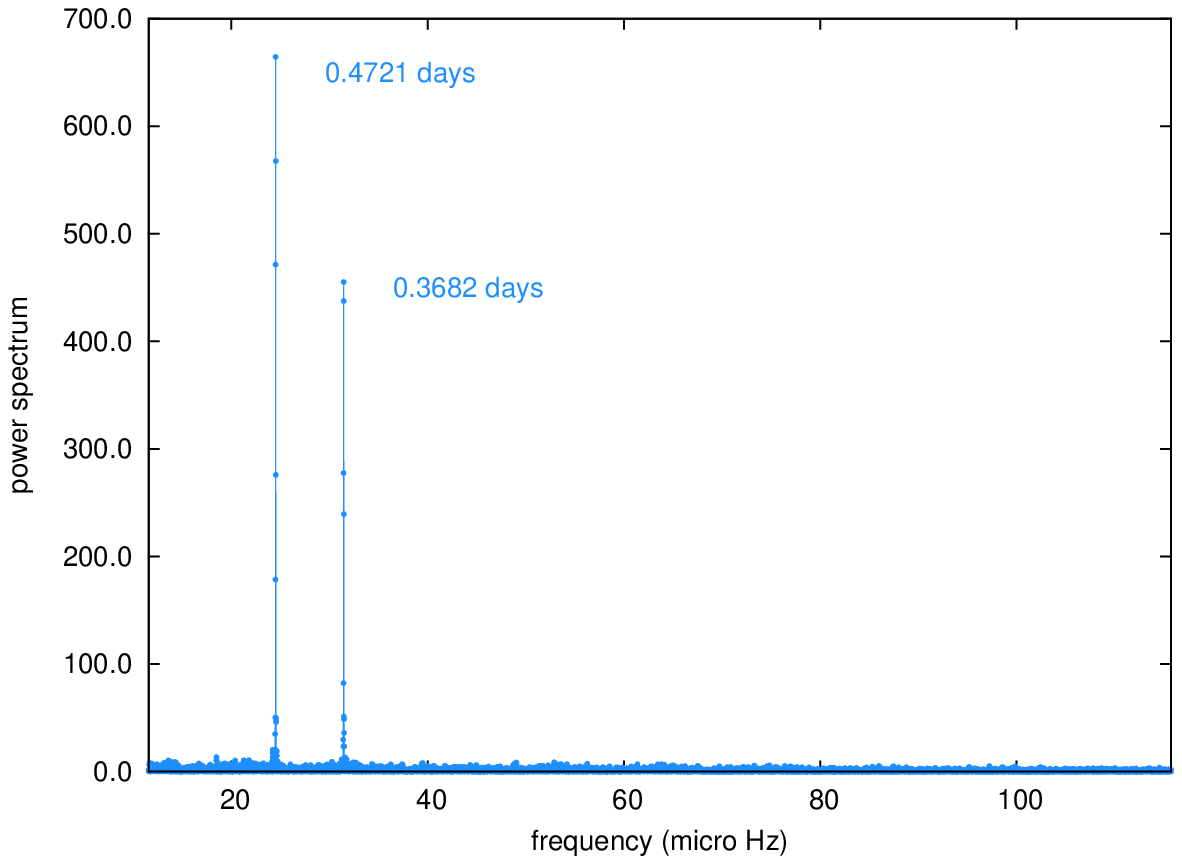}
    \caption{Lomb-Scargle periodogram of KIC 6751111 (dM4.5).  Two persistent periodicities are present in the data.}
 \end{figure}     

\begin{figure}[t!]
    \label{KIC6751111_Q9}
    \centering
    \includegraphics[width=12.5cm]{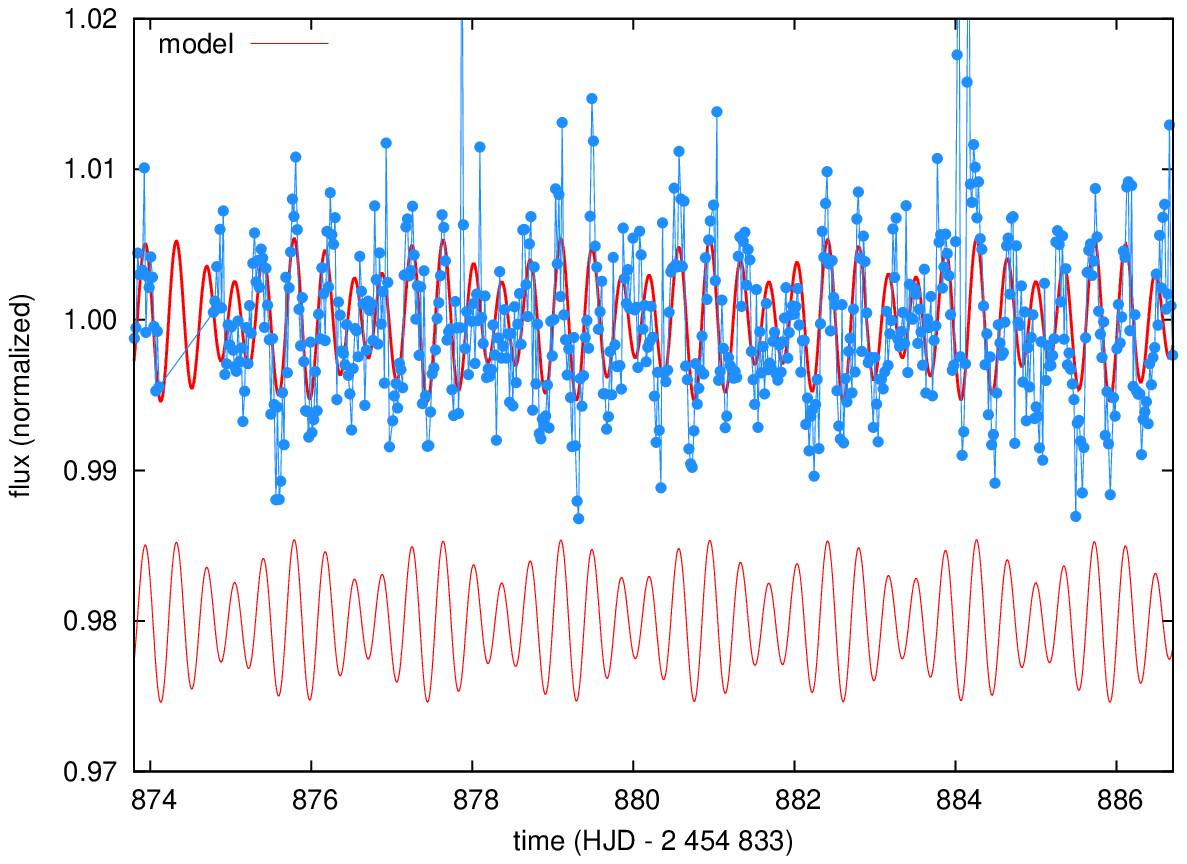}
    \caption{Filtered Kepler light curve in Quarter 9 (upper line with dots) for KIC 6751111 compared with our model using the two periodicities found in the periodogram (lower curve). The model light curve is also superimposed on the data to show the fit.} 
 \end{figure}   
  
Ten targets have clear dominant periods shorter than 7 days that we interpret as due to rotational modulation of long lived surface 
features. The frequency of fast rotators in our sample (vrot$\ge$3 km/s corresponding to Prot shorter than about 3 days) 
is 55\% , which is in good agreement with previous estimates based on spectroscopic measurements of rotational broadening (\cite{2012AJ...144.99D}) that find about 65\% of late dMs with vsin i higher than 12 km/s. 

The high precision of Kepler allows us to obtain reliable rotational periods in 61\% of the sample, a much higher 
detection rate than in ground based transit surveys like MEarth, which reported rotation periods for 41 out of 273 (i.e., 15\%) fully 
convective M dwarfs (\cite{2011ApJ.727.56I}),  but comparable to the Kepler based analysis for M dwarfs reported by \cite{2013arXiv1303.6787M}, which detected rotation periods in 63\% of their sample (only one object in common with our sample).

The pattern of rotational variability often changes from one quarter to another, indicating that the surface features evolve with time. An example of this behaviour is shown in Figure 10.  More data is needed to investigate the possible cyclic recurrence of these patterns. 

\begin{figure}[t!]
    \label{KIC8869922}
    \centering
    \includegraphics[width=12.5cm]{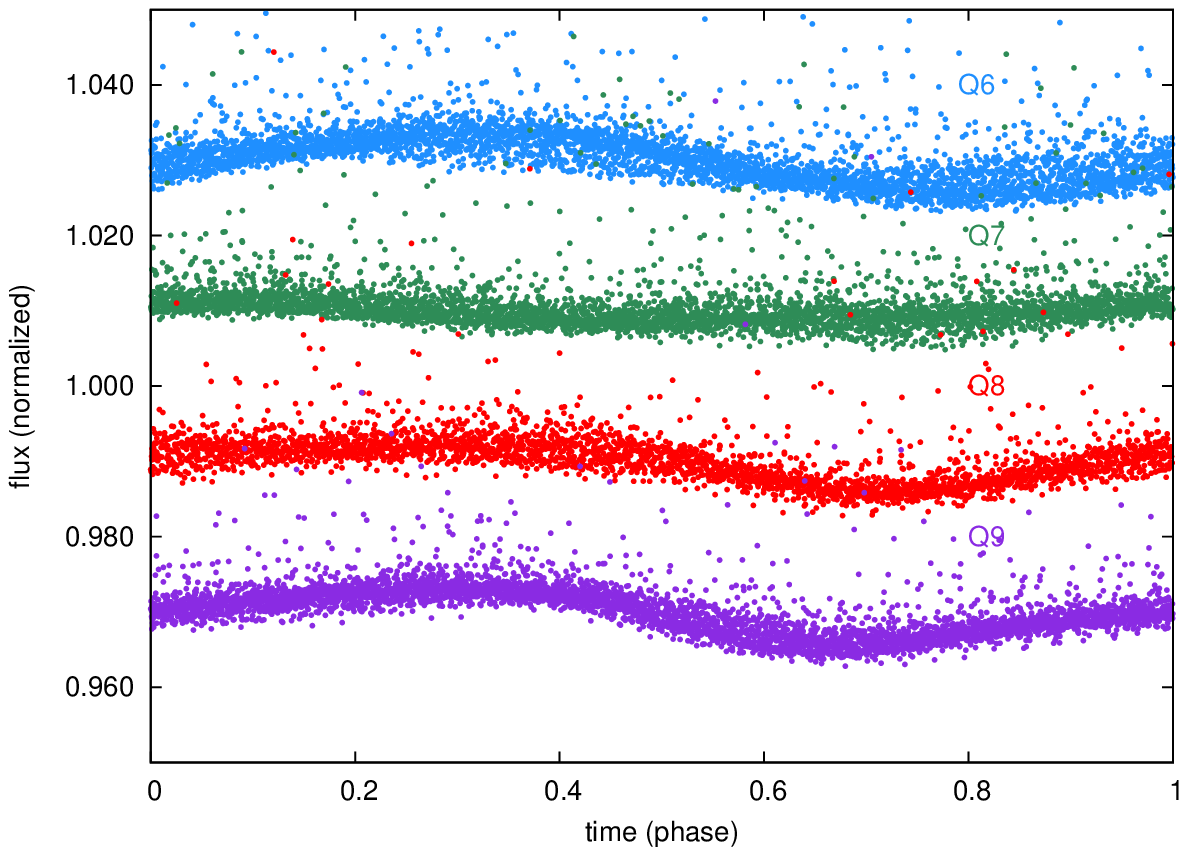}
    \caption{Kepler light curves of KIC 8869922 folded with the same rotation period of 0.70 days in four different quarters (coded with different colors) to illustrate the evolution of surface features. The frequent upward excursions in flux are due to flares.}
 \end{figure}   
  
The periods marked with colons in Table 3 are uncertain, and correspond to quiet VLM stars with possible 
rotation periods longer than 30 days. An example of our sinusoidal fit to the light curve of a slow rotator is given in Figure 11. Such long 
periods are not uncommon among VLM stars (\cite{2011ApJ.727.56I}). Even longer photometric rotation periods (80-130 days) have been reported from Hubble Space Telescope monitoring of the very nearby VLM stars Proxima Centauri and Barnard`s star (\cite{1998AJ...116.429B}). However, we consider our long periods as tentative because we cannot rule out that they may arise from uncorrected long-term effects in the Kepler data. 

\begin{figure}[t!]
    \label{KIC7799941}
    \centering
    \includegraphics[width=12.5cm]{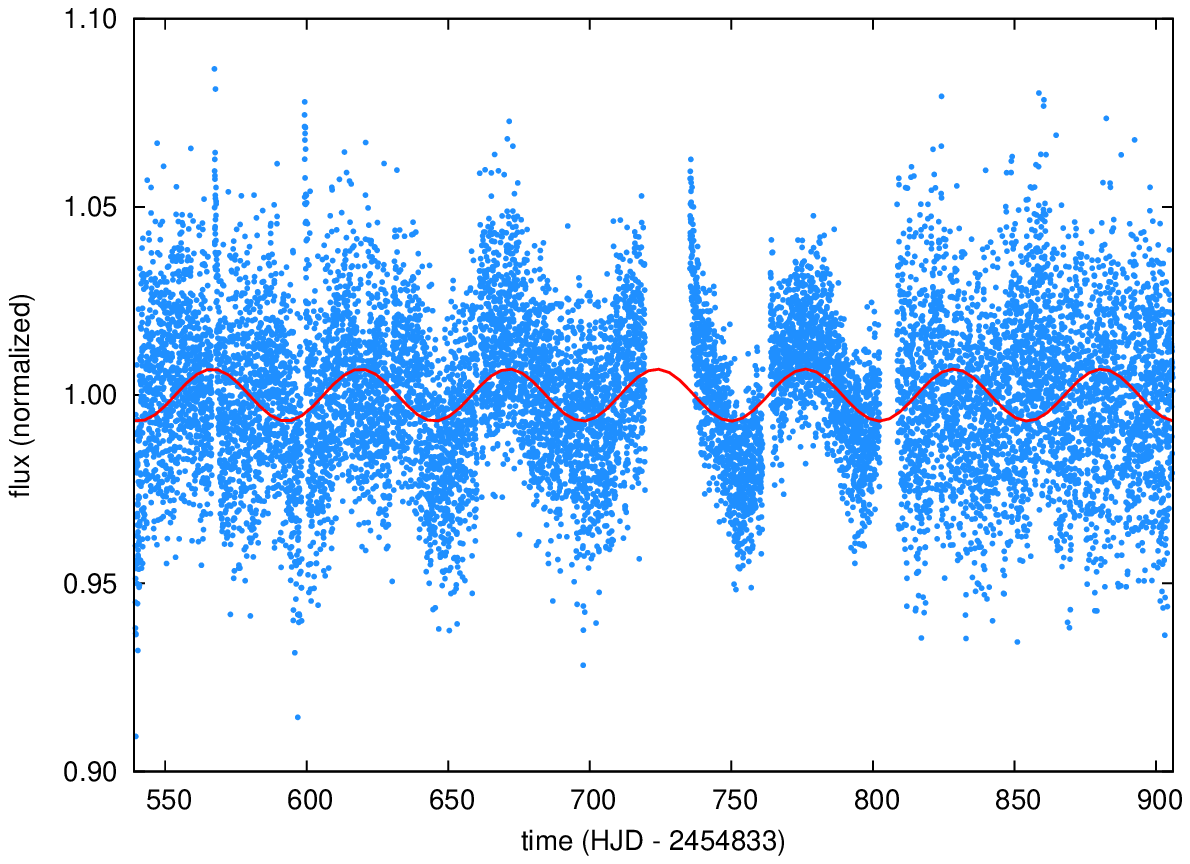}
    \caption{The Kepler light curve (dots) of KIC 7799941 compared with our sinusoidal model (continuous line) with a rotation period of 
53 days.}
 \end{figure}

\subsubsection{Flares}

 We searched for flares using a custom made algorithm to identify them. The algorithm proceeds in the following steps: (a) it calculates the median value of the light curve in 2 day bins; (b) it calculates the median deviation; (c) it searches for upward flux excursions larger than a threshold (the threshold was set at a value equal to the median plus 5 times the median deviation); (d) it identifies as a flare events where the number of points above the threshold is larger or equal to 3.  This is a conservative criterion to elliminate high frequency noise, intrumental glitches or cosmic rays. 
The number of flares detected in each target are given in column 4 of Table 3. Figures 12 and 13 display the light curves and the thresholds used to detect the flares. Each flare detected is marked with a circle at the peak of the flux. 

\begin{figure}[t!]
    \label{FlareDet}
    \centering
    \includegraphics[width=9.9cm,angle=270]{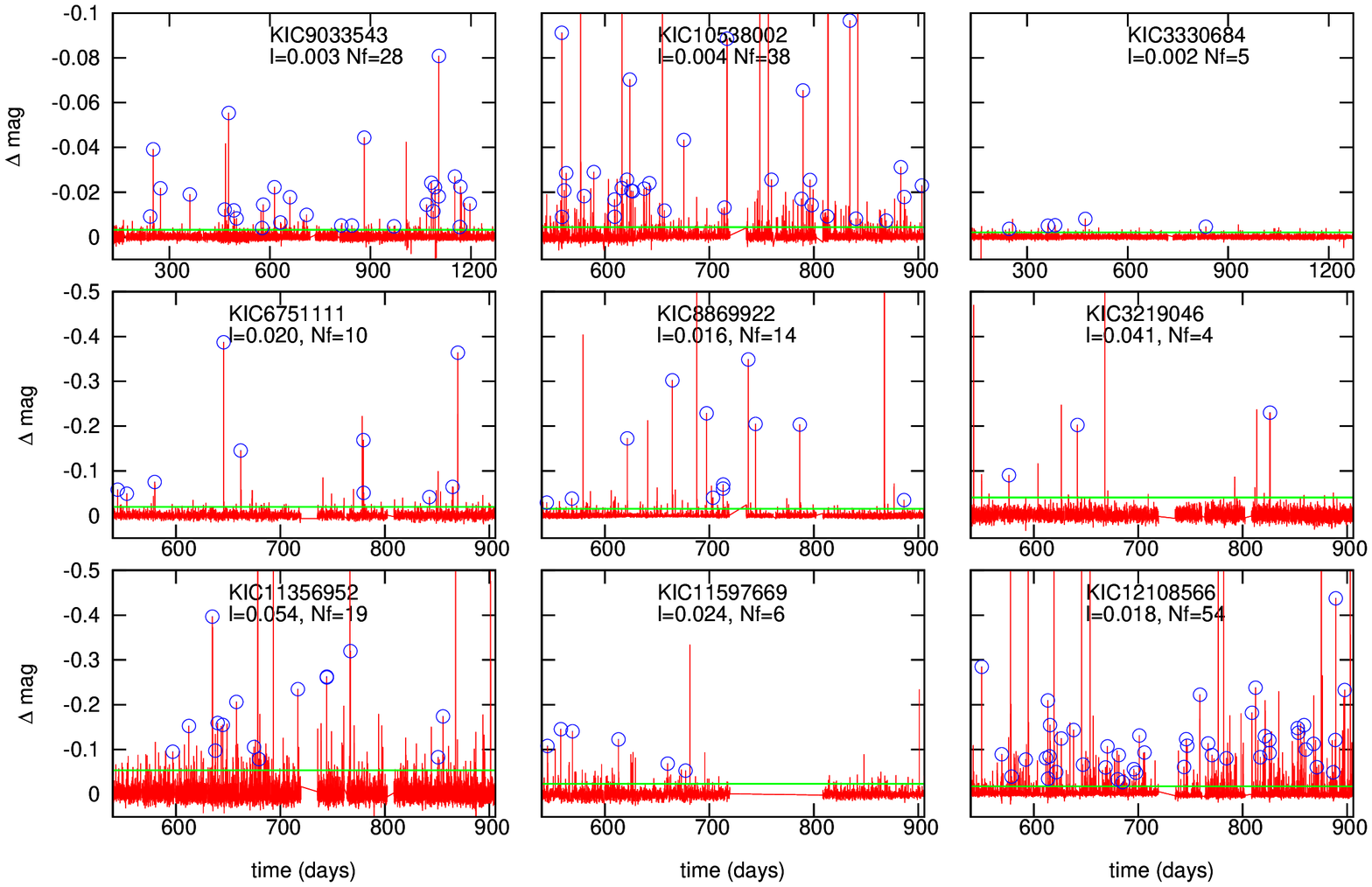}
    \caption{A mosaic of our lightcurves showing more than 3 flares and the thresholds used to detect them (horizontal green lines). The number of flares (Nf) is given for each target.}
 \end{figure}

\begin{figure}[t!]
    \label{FlareDet}
    \centering
    \includegraphics[width=9.9cm,angle=270]{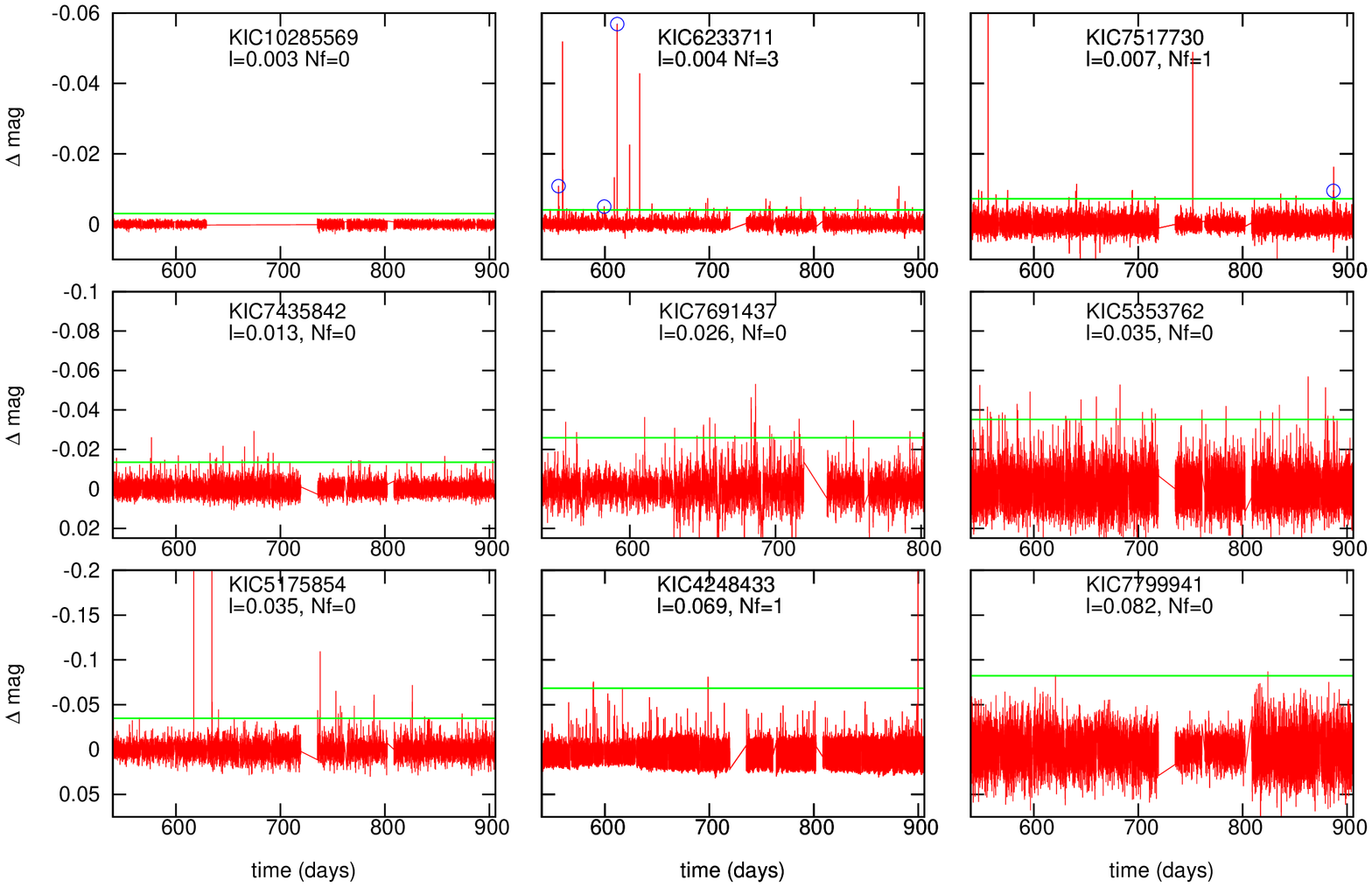}
    \caption{A mosaic of our lightcurves showing less than 3 flares and the thresholds used to detect them (horizontal green lines). The number of flares (Nf) is given for each target.}
 \end{figure}   
  
Very large flare events where the flux of the star surpassed in more than half a magnitude the quiescent brightness 
were seen in KIC 8869922 (dM5), KIC 3219406 (dM4.5), KIC 4248433 (dM5), KIC 5175854 (dM5), KIC 12108566 (dM5), 
KIC 10538002 (dM5), KIC 11356952 (dM8.5). The largest of them all was seen in KIC 12108566. It reached a peak flux 9 times larger 
than the quiescent level and lasted several hours.  

Flare rates were defined as the number of pixels were flares had been detected divided by the number of useful points in the 
light curve.  The flare rates provided by us are likely lower limits in most cases because some short flares may have been missed by our conservative algorithm. A caveat of our method is that the flare rate obtained depends on the noise characteristics of the light curves.

The rate of flares calculated for each target is given in Table 3, and Figure 14 plots flare rates versus  H$_\alpha$ equivalent widths. VLM stars with stronger H$_\alpha$ in emission have a higher incidence of flares, as expected because both phenomena are thought to be connected to the strength of magnetic fields. Three out of five VLM stars with H$_\alpha$ equivalent width 
$\le$ 1 \AA , do not show any flare detection, and the other two have low flare rates.

\begin{figure}[t!]
    \label{FlaresHal}
    \centering
    \includegraphics[width=12.5cm]{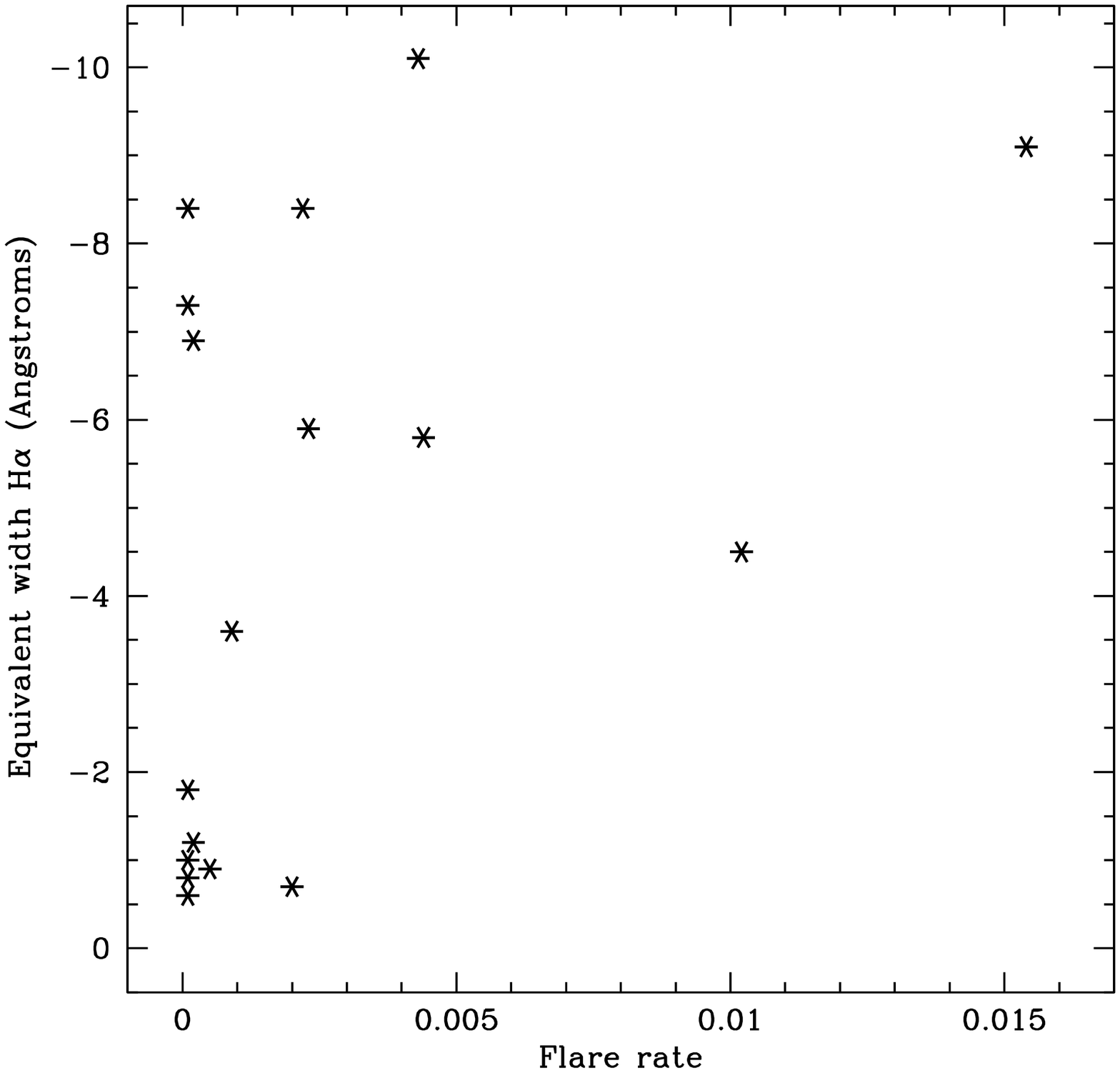}
    \caption{H$_\alpha$ equivalent widths versus flare rates for our sample.}
 \end{figure}

\section{Stellar parameters}

So far direct radius measurements obtained via ground-based interferometric observations are available for only two VLM stars (e.g., \cite{2012ApJ....757.112B}, and references therein), namely, GJ 699 (dM4) and GJ 551 (dM5.5). 
A theoretical mass-spectral type relationship down to the substellar limit has been constructed by \cite{1996ApJ....461.L51B}, but 
the luminosities that they used are not consistent with the interferometric data. For the sake of estimating the detectability of planetary transits 
in the Kepler light curves of our targets we have used a linear relationship for the two VLM stars cited above and we have interpolated 
or extrapolated it using the spectral subclasses assigned to our targets. At spectral types later than dM7 we assume that the radius 
is constant due to electron degeneracy support. The results are provided in Table 4. 

Effective temperatures ($T_\mathrm{eff}$)  and surface gravities (log g) for our targets were obtained from $\chi^{2}$ fitting of the spectral energy distribution (SED)  using VOSA\footnote{http://svo.cab.inta-csic.es/theory/vosa/} (Virtual Observatory SED Analyzer, \cite{2008A&A...492..277B}). 
Figure 15 provides an example of VOSA SED fitting. 
VOSA is a VO-tool designed to query several photometric catalogs accessible through VO services as well as VO-compliant theoretical models and perform an statistical test to determine which model reproduces best the observed data. This approach 
has already been used to reveal new BDs in the WISE survey \cite{2011A&A...534..L7A}. VOSA was able to distinguish the giant in our sample, 
and it generally provides $T_\mathrm{eff}$ that agree with our optically determined spectral types and the relation between $T_\mathrm{eff}$ 
and spectral subclass given by \cite{2004AJ....127.3516G}.  The atmosphere models used by VOSA were the BT-Settl (\cite{2012EAS....57.3A}).

VOSA provided satisfactory fits for all our targets except for  KIC 7691437. 
Archive images were inspected visually and it was found that KIC 7691437 WISE images are blended and hence the VOSA results are not 
reliable. No blending with other objects was noticed for any of the other targets. For KIC 7691437 the $T_\mathrm{eff}$ was obtained 
directly from the spectral type using the relation in \cite{2004AJ....127.3516G}.

\begin{table*}
\centering
\caption{Astrophysical parameters.}
\label{tabla4}
\scriptsize
\hspace{-1cm}
\begin{tabular}{ccccccc}
\noalign{\smallskip}
\hline
\noalign{\smallskip}
KIC	& SpT & Teff   &  log g     & M/H & Radius   & Av      \\
 	&        &  K     &  cm/s2   &          &  $R/R_{\odot}$    &           \\
\noalign{\smallskip}
\hline
\noalign{\smallskip}
6751111   &  dM4.5  &  3000$\pm$50       &  5.0$\pm$0.5  &  0.0$\pm$0.5  & 0.17$\pm$0.01    &   0.1$\pm$0.1   \\ 
7799941   &  dM7     &  2700$\pm$300     &  5.5$\pm$0.7  &  0.5$\pm$0.5  & 0.10$\pm$0.01    &   0.4$\pm$0.1   \\
11597669 &  dM5     &  3150$\pm$50       &  5.0$\pm$0.5  &   0.3$\pm$0.5 & 0.15$\pm$0.01    &   0.0        \\
8869922   &  dM5     &  3100$\pm$50       &  5.0$\pm$0.5  &   0.3$\pm$0.5  & 0.16$\pm$0.01   &   0.1$\pm$0.1        \\ 
3219046   &  dM4.5  &  3000$\pm$50       &  5.5$\pm$0.5  &  -0.5$\pm$0.5 & 0.17$\pm$0.01   &  0.0         \\ 
4248433   &  dM5     &  3000$\pm$50       &  4.5$\pm$0.5  &   0.0$\pm$1.0  & 0.16$\pm$0.01   &   0.0        \\ 
3330684   &  dM5     &  2900$\pm$50       &  4.5$\pm$0.5  &   0.0$\pm$0.5  & 0.16$\pm$0.01   &   0.0        \\  
5175854   &  dM5     &  2700$\pm$100     &  4.5$\pm$0.5  &   0.5$\pm$0.5  & 0.16$\pm$0.01   &   0.1$\pm$0.1        \\  
7517730   &  dM6     &  2800$\pm$50       &  4.0$\pm$0.5  &   0.0$\pm$0.5  & 0.13$\pm$0.01    &   0.1$\pm$0.1        \\ 
7435842   &  dM6     &  2700$\pm$200     &  5.0$\pm$1.0  &   0.5$\pm$0.5   & 0.13$\pm$0.01   &   0.0                \\ 
5353762   &  dM6.5     &  2700$\pm$200     &  4.5$\pm$1.0  &   0.5$\pm$0.5   & 0.12$\pm$0.01   &   0.0                 \\ 
12108566 &  dM5     &  3000$\pm$200     &  5.5$\pm$0.5  &   0.3$\pm$0.5   & 0.16$\pm$0.01   &  0.1$\pm$0.1  \\ 
10538002 &  dM5     &  3000$\pm$50       &  4.5$\pm$0.5  &   0.0$\pm$0.5   & 0.16$\pm$0.01   &    0.0               \\ 
7691437   &  dM7.5  &  2500$\pm$150     &                         &                          & 0.10$\pm$0.01    &                        \\ 
11356952 &  dM8.5  &  2600$\pm$300     &  5.0$\pm$0.5   &   0.5$\pm$0.5  & 0.08$\pm$0.01    &  0.0                  \\ 
9033543   &  dM5.5     &  2900$\pm$50       &  5.0$\pm$0.5   &   0.3$\pm$0.5  & 0.14$\pm$0.01    &  0.2$\pm$0.1   \\ 
10285569 &  dM4.5  &  3150$\pm$50       &  4.5$\pm$0.5   & -0.3$\pm$0.5  & 0.17$\pm$0.01    &   0.1$\pm$0.1   \\ 
8450707   &  M4 III   &  3350$\pm$50       &  -0.5$\pm$0.5 &   0.0                  &                             &   1.0$\pm$0.1    \\ 
6233711   &  dM4.5  &  3200$\pm$50       &  5.0$\pm$0.5   &   0.5$\pm$0.5  & 0.17$\pm$0.01    &   0.0                   \\ 
\noalign{\smallskip}
\hline
\noalign{\smallskip}
\noalign{\smallskip}
\hline
\noalign{\smallskip}
\end{tabular}
\end{table*}  
 
\begin{figure}[t!]
    \label{VOSA}
    \centering
    \includegraphics[width=13.5cm]{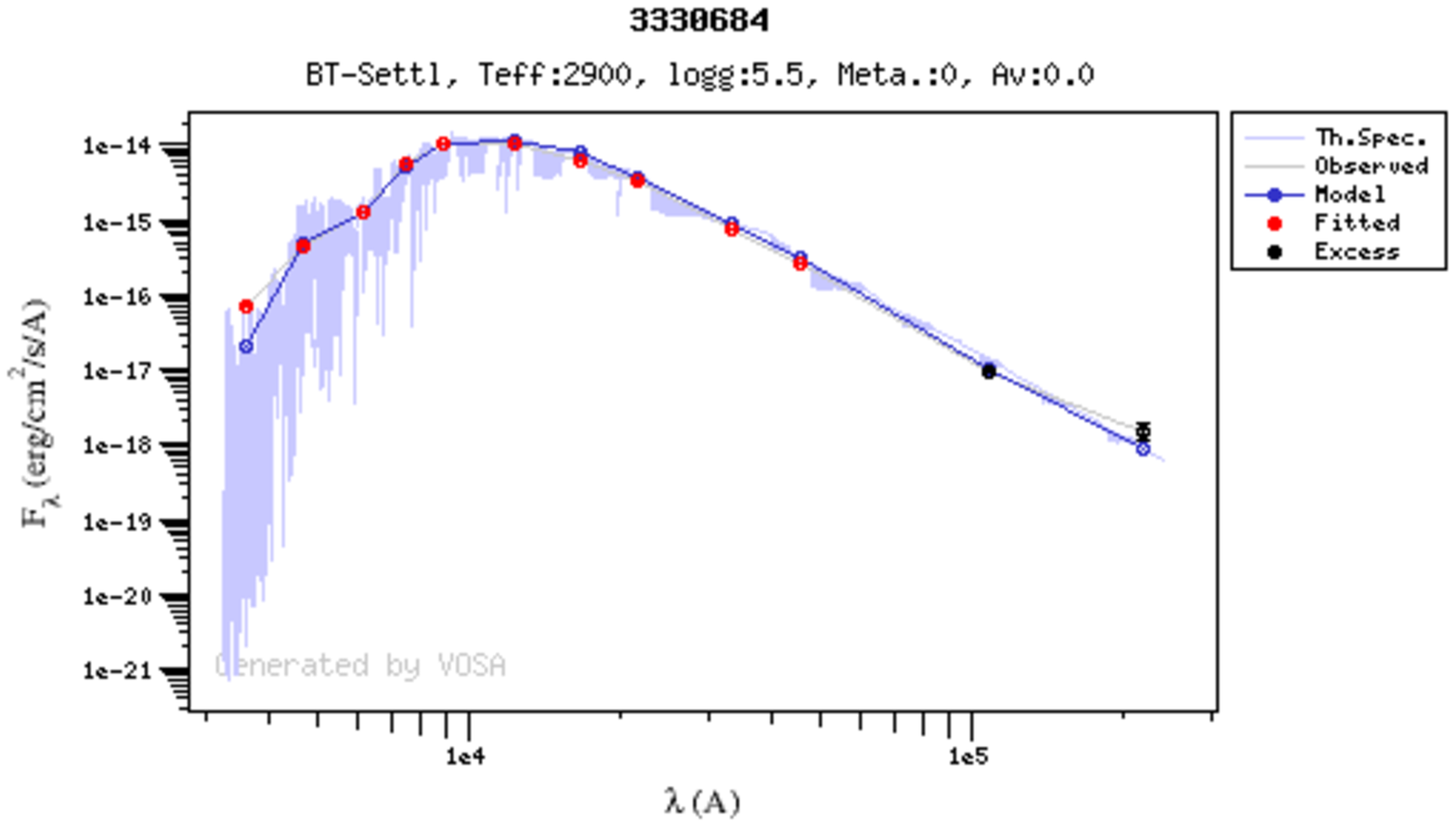}
    \caption{VOSA SED fitting for one of our confirmed KIC VLM dwarfs (KIC 3330684, dM5). Catalogue and synthetic photometric points are represented in red and blue, respectively. Points not considered in the fitting are shown in black. The best fitting theoretical spectrum (from which the atmospheric parameters were derived) is also plotted.}
 \end{figure}     

\subsection{Transversal velocities}

Spectrophotometric distances for the confirmed VLM dwarfs were calculated using the 2MASS K-band magnitudes and the spectral type versus absolute K-band magnitude provided by \cite{2002ApJ....564.452L}. Transversal velocities were obtained from the total proper motion and the distance in parsecs. Results are provided in Table 5. 

\begin{table*}
\centering
\caption{Spectrophotometric distances and transversal velocities of KIC VLM dwarfs.}
\label{tabla5}
\scriptsize
\hspace{-1cm}
\begin{tabular}{ccccccc}
\noalign{\smallskip}
\hline
\noalign{\smallskip}
KIC	& SpT & K$_{2MASS}$  & M$_K$ & m-M            &  D     & Vt        \\
 	&        &                       &             & $\pm$0.08  & pc    &  km/s         \\
\noalign{\smallskip}
\hline
\noalign{\smallskip}
6751111   &  dM4.5  &  12.39  & 8.18  & 4.21  & 69.5$\pm$2.6  &  33.3$\pm$1.2    \\ 
7799941   &  dM7     &  12.86  & 9.90  & 2.96  & 39.1$\pm$1.4  &  21.5$\pm$0.8    \\ 
11597669 &  dM5     &  11.89  & 8.41  & 3.48  & 49.7$\pm$1.8  &  27.8$\pm$1.0    \\  
8869922   &  dM5     &  11.02  & 8.41  & 2.61  & 33.3$\pm$1.2  &  21.1$\pm$0.7    \\
3219046   &  dM4.5  &  12.62  & 8.18  & 4.44  & 77.3$\pm$2.8  &  44.0$\pm$2.1    \\ 
4248433   &  dM5     &  10.44  & 8.41  & 2.03  & 25.5$\pm$0.9  &  15.9$\pm$0.6    \\
3330684   &  dM5     &  10.31  & 8.41  & 1.90  & 24.0$\pm$0.9  &  37.5$\pm$1.4    \\  
5175854   &  dM5     &  12.93  & 8.41  & 4.52  & 80.2$\pm$3.0  &  66.1$\pm$2.5    \\  
7517730   &  dM6     &  11.54  & 9.17  & 2.37  & 29.8$\pm$1.1  &  26.3$\pm$1.0    \\
7435842   &  dM6     &  12.08  & 9.17  & 2.91  & 38.2$\pm$1.4  &                             \\
5353762   &  dM6.5     &  13.06  & 9.17  & 3.89  & 60.0$\pm$2.2  &  48.9$\pm$1.8    \\ 
12108566 &  dM5     &  12.34  & 8.41  & 3.93  & 61.1$\pm$2.3  &  41.4$\pm$1.6    \\ 
10538002 &  dM5     &  10.63  & 8.41  & 2.22  & 27.8$\pm$1.0  &  22.3$\pm$0.8    \\ 
7691437   &  dM7.5  &  11.79  & 9.92  & 1.87  & 23.7$\pm$0.9  &  31.9$\pm$1.1    \\ 
11356952 &  dM8.5  &  12.60  & 10.55  & 2.05  & 25.7$\pm$0.9  &  16.9$\pm$0.7    \\ 
9033543   &  dM5.5     &  10.38  & 9.17  & 1.21  & 17.5$\pm$0.6  &  25.1$\pm$0.9    \\ 
10285569 &  dM4.5  &  11.00  & 8.18  & 2.82  & 36.6$\pm$1.3  &  55.5$\pm$2.0    \\ 
6233711   &  dM4.5  &  11.52  & 8.18  & 3.34  & 46.5$\pm$1.7  &  27.5$\pm$1.1    \\ 
\noalign{\smallskip}
\hline
\noalign{\smallskip}
\noalign{\smallskip}
\hline
\noalign{\smallskip}
\end{tabular}
\end{table*}

Transversal velocities have been used as a proxy for dynamical age in VLM stars. \cite{2007ApJ....666.1205Z} found that the mean transversal 
velocity of L and T dwarfs in the solar vicinity is lower than that of solar-type stars, and suggested that the UCDs could be kinematically 
younger, but a study of a larger sample that included late-M dwarfs did not reach the same conclusions (\cite{2009AJ...137.1F}). We have calculated equatorial rotational velocities for our targets using the rotational periods given in Table 3 and radii in Table 5. The rotational velocities are plotted versus the transversal velocities in Figure 16. There seems to be a trend towards lower rotational velocity for higher transversal velocity as might be expected from evolutionary effects as the stars get older. This trend requires confirmation with a larger sample and with more accurate proper motion values and radial velocity measurements for the VLM stars observed by Kepler. 

\begin{figure}[t!]
    \label{VtanVrot}
    \centering
    \includegraphics[width=9.5cm, angle=270]{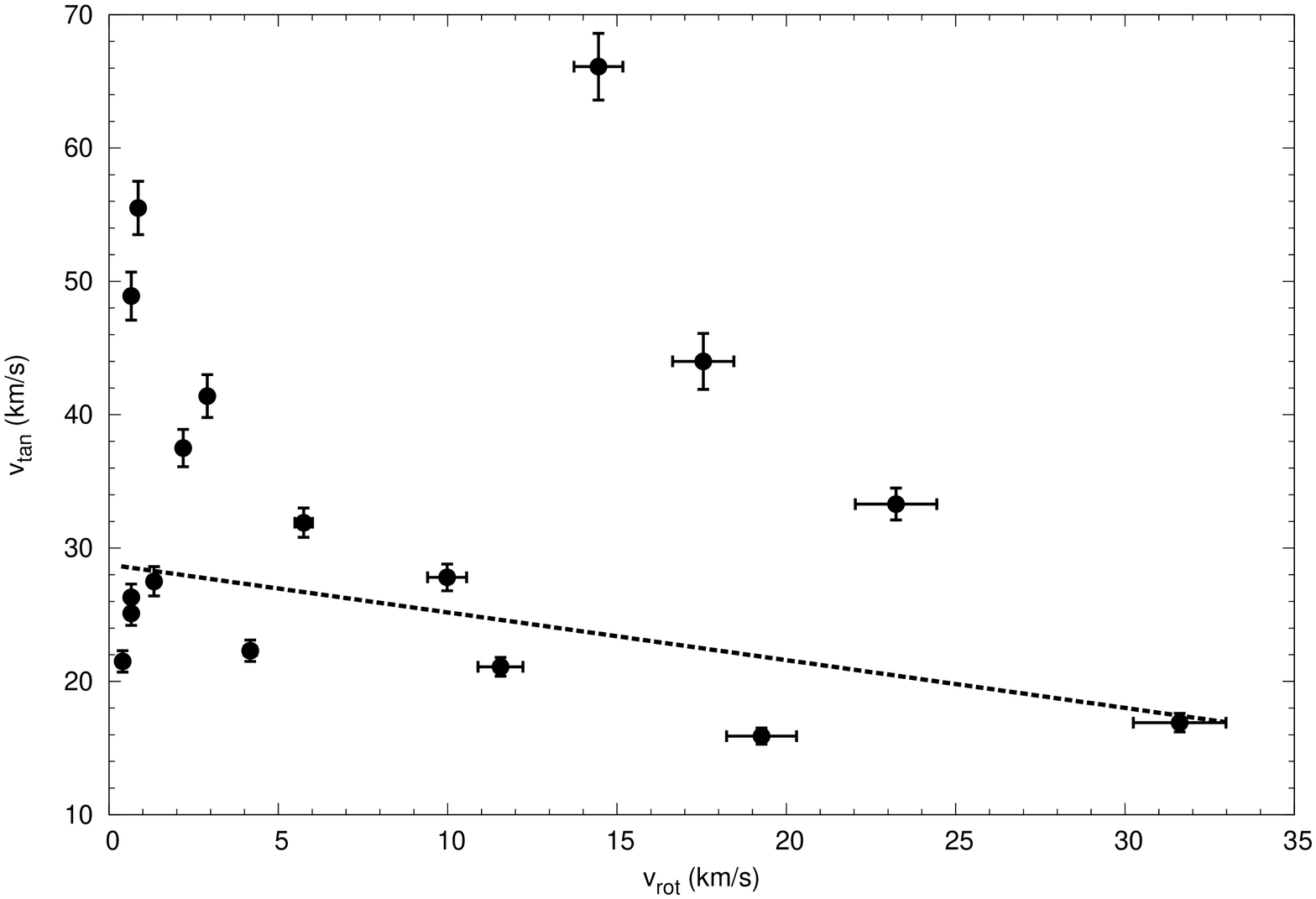}
    \caption{Transversal velocities versus equatorial rotational velocities for 17 KIC VLM stars.  A linear fit regression to the data is  shown 
as a dotted line. All velocities are measured in units of km/s.}
 \end{figure}

\section{Planet detectability}

Planet detectability depends on the magnetic activity of the star, which is also related to the stellar rotation. Cool stars show higher 
level of chromospheric activity for faster rotation, and also higher jitter in radial velocity (\cite{1998ApJL....498.L153S}) and photometric flux variability (\cite{2011AA....533.A44L}).  Thus, planet detectability tends to be reduced in active cool stars, but it is not clear whether or not this trend continues into the ultracool dwarf domain. 

\begin{figure}[t!]
    \label{SizesHal}
    \centering
    \includegraphics[width=12.5cm]{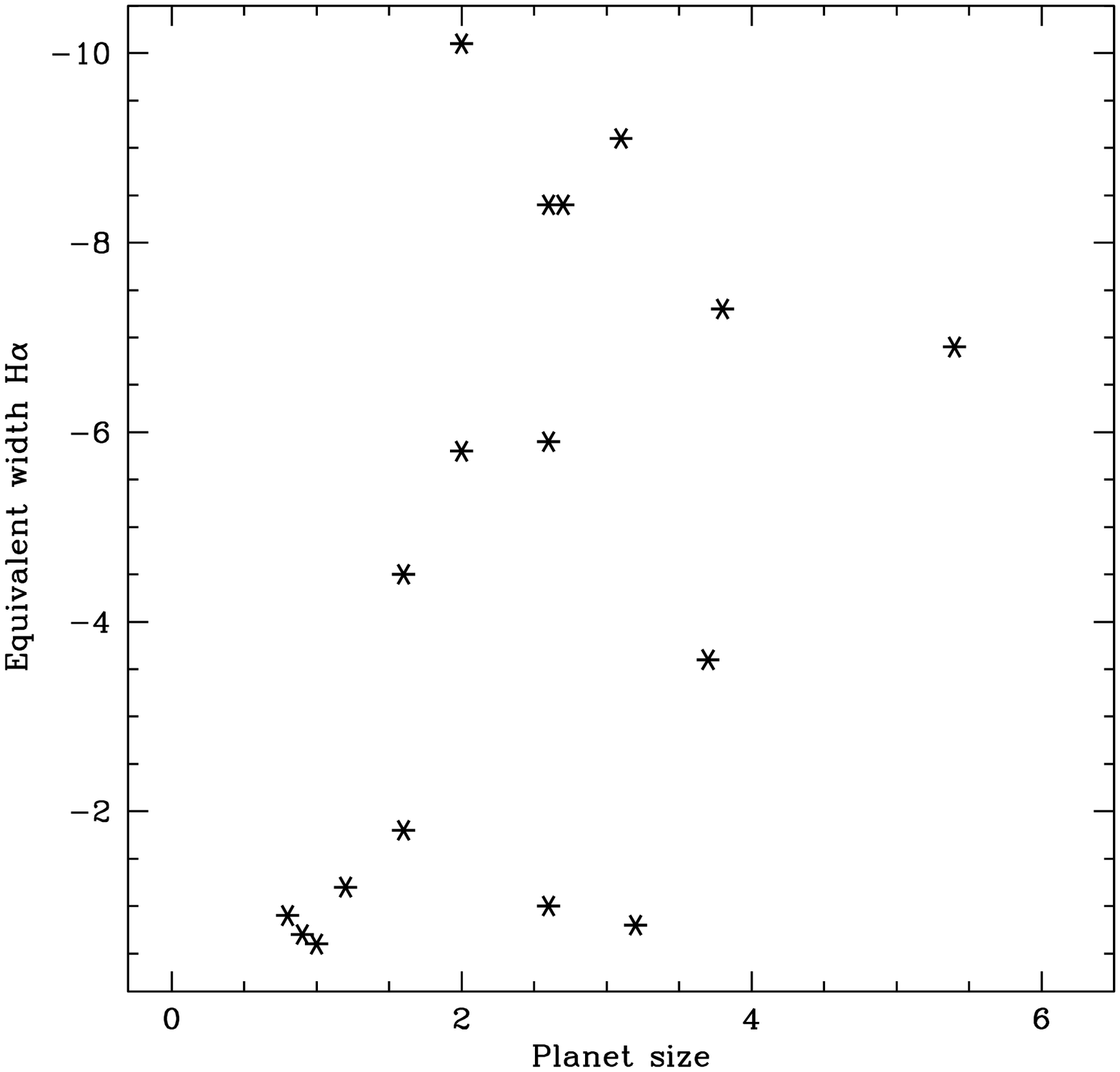}
    \caption{H$_\alpha$ equivalent widths versus planet size detectability for our sample. Equivalent widths in \AA , as listed in Table 2, and planet sizes in Earth radii as listed in Table 3.}
 \end{figure}   
  
In our sample, we have checked the relation between H$_\alpha$ equivalent width as a proxy of chromosperic activity and planet size limits as 
a proxy to planet detectability. The results are shown in Figure 17. Super-Earth planets with sizes around 2 Earth radii are detectable for any level of chromospheric activity. However, the detectability of the lowest planet sizes, around and even slightly below the radius of the Earth, is reached only for the quietest VLM stars. The result that the best sensitivity for planetary transits in VLM stars with Kepler is reached for the least chromospherically active VLM stars is tentative and needs to be tested with a larger sample.   

\begin{acknowledgements}
This work is based (in part) on observations made with the Gran Telescopio Canarias (GTC), operated on the island of La Palma in the Spanish Observatorio del Roque de los Muchachos of the Instituto de Astrof\'{\i}sica de Canarias. This research has been supported by the Spanish Ministry of Economy and Competitiveness (MINECO) under the grant AyA2011-30147-C03-03 and by the 
Consolider Ingenio 2010-GTC project. 
Part of the manuscript was written while EM was a visiting professor in the Geosciences Department at the 
University of Florida. 
The data presented here were obtained (in part) with ALFOSC, which is provided by the Instituto de Astrof\'{\i}sica de Andalucia (IAA) under a joint agreement with the University of Copenhagen and the NBIfAFG of the Astronomical Observatory of Copenhagen. This article is partly based on observations obtained with the Nordic Optical Telescope, operated on the island of La Palma jointly by Denmark, Finland, Iceland, Norway, and Sweden, in the Spanish Observatorio del Roque de los Muchachos of the Instituto de Astrof\'{\i}sica de Canarias. 
This publication makes use of data products from the Two Micron All Sky Survey (2MASS), which is a joint project of the University of Massachusetts and the Infrared Processing and Analysis Center/California Institute of Technology, funded by the National Aeronautics and Space Administration and the National Science Foundation.
Funding for the SDSS and SDSS-II has been provided by the Alfred P. Sloan Foundation, the Participating Institutions, the National Science Foundation, the U.S. Department of Energy, the National Aeronautics and Space Administration, the Japanese Monbukagakusho, the Max Planck Society, and the Higher Education Funding Council for England. The SDSS Web Site is http://www.sdss.org/. The SDSS is managed by the Astrophysical Research Consortium for the Participating Institutions. The Participating Institutions are the American Museum of Natural History, Astrophysical Institute Potsdam, University of Basel, University of Cambridge, Case Western Reserve University, University of Chicago, Drexel University, Fermilab, the Institute for Advanced Study, the Japan Participation Group, Johns Hopkins University, the Joint Institute for Nuclear Astrophysics, the Kavli Institute for Particle Astrophysics and Cosmology, the Korean Scientist Group, the Chinese Academy of Sciences (LAMOST), Los Alamos National Laboratory, the Max-Planck-Institute for Astronomy (MPIA), the Max-Planck-Institute for Astrophysics (MPA), New Mexico State University, Ohio State University, University of Pittsburgh, University of Portsmouth, Princeton University, the United States Naval Observatory, and the University of Washington. 
This research has made use of the NASA Astrophysics Data System and of SIMBAD operated by the Centre de Don{\'e}es Astronomiques de Strasbourg.
IRAF is distributed by the National Optical Astronomy Observatory, which is operated by the Association of Universities for Research in Astronomy (AURA) under cooperative agreement with the National Science Foundation. 
Atmospheric parameters were estimated using VOSA, a VO-tool developed under the Spanish Virtual Observatory project supported from the Spanish MICINN through grant AyA2008-02156. The detailed referee report provided by Dr. Alex Scholz was useful to improve the original manuscript.

\end{acknowledgements}

\end{document}